\documentclass[fleqn,usenatbib]{mnras}

\usepackage{newtxtext,newtxmath}

\usepackage[T1]{fontenc}

\DeclareRobustCommand{\VAN}[3]{#2}
\let\VANthebibliography\thebibliography
\def\thebibliography{\DeclareRobustCommand{\VAN}[3]{##3}\VANthebibliography}

\usepackage{graphicx}	
\usepackage{amsmath}	
\usepackage{multirow}
\usepackage{caption}
\usepackage{float}
\usepackage{orcidlink}

\defcitealias{Tycho2}{Tycho-2}
\defcitealias{2MASS}{2MASS}
\defcitealias{GaiaDR3}{\textit{Gaia} DR3}
\defcitealias{Wright2010}{\textit{WISE}}

\title[The sub-Neptune orbiting HD 77946]{Confronting compositional confusion through the characterisation of the sub-Neptune orbiting HD 77946}

\author[L. Palethorpe et al.]{L. Palethorpe\orcidlink{0000-0002-1664-4105},$^{1,2}$\thanks{E-mail: larissa.palethorpe@ed.ac.uk} A. Anna John\orcidlink{0000-0002-1715-6939},$^{3,4}$ A. Mortier\orcidlink{0000-0001-7254-4363},$^{5}$ J. Davoult\orcidlink{0000-0002-6177-2085},$^{6}$ T. G. Wilson\orcidlink{0000-0001-8749-1962},$^{3,4}$ K. Rice\orcidlink{0000-0002-6379-9185},$^{1,2}$ \newauthor A. C. Cameron\orcidlink{0000-0002-8863-7828},$^{3,4}$ Y. Alibert\orcidlink{0000-0002-4644-8818},$^{6,7}$ L. A. Buchhave\orcidlink{0000-0003-1605-5666},$^{8}$ L. Malavolta\orcidlink{0000-0002-6492-2085},$^{9}$ J. Cadman\orcidlink{0000-0002-3200-3121},$^{1,2}$ M. López-Morales\orcidlink{0000-0003-3204-8183},$^{10}$ \newauthor  X. Dumusque\orcidlink{0000-0002-9332-2011},$^{11}$ A. M. Silva\orcidlink{0000-0003-4920-738X},$^{12,13}$ S. N. Quinn\orcidlink{0000-0002-8964-8377},$^{10}$ V. Van Eylen\orcidlink{0000-0001-5542-8870},$^{14}$ S. Vissapragada\orcidlink{0000-0003-2527-1475},$^{10}$ L. Affer\orcidlink{0000-0001-5600-3778},$^{15}$ \newauthor D. Charbonneau\orcidlink{0000-0002-9003-484X},$^{10}$ R. Cosentino\orcidlink{0000-0003-1784-1431},$^{16}$ A. Ghedina\orcidlink{0000-0003-4702-5152},$^{16}$ R. D. Haywood\orcidlink{0000-0001-9140-3574},$^{17,18}$ D. W. Latham\orcidlink{0000-0001-9911-7388},$^{10}$ \newauthor F. Lienhard\orcidlink{0000-0003-4047-0771},$^{19}$ A. F. Martínez Fiorenzano\orcidlink{0000-0002-4272-4272},$^{16}$ M. Pedani\orcidlink{0000-0002-5752-6260},$^{16}$ F. Pepe,$^{11}$ M. Pinamonti\orcidlink{0000-0002-4445-1845},$^{20}$ A. Sozzetti\orcidlink{0000-0002-7504-265X},$^{20}$ \newauthor M. Stalport\orcidlink{0000-0003-0996-6402},$^{12}$ S. Udry\orcidlink{0000-0001-7576-6236},$^{13}$ and A. Vanderburg\orcidlink{0000-0001-7246-5438}$^{21}$
\\
$^{1}$Institute for Astronomy, University of Edinburgh, Royal Observatory, Blackford Hill, Edinburgh, EH9 3HJ, UK\\
$^{2}$Centre for Exoplanet Science, University of Edinburgh, Edinburgh, EH9 3HJ, UK\\
$^{3}$SUPA School of Physics and Astronomy, University of St Andrews, North Haugh, St Andrews, KY16 9SS, UK\\
$^{4}$Centre for Exoplanet Science, University of St Andrews, North Haugh, St Andrews, KY169SS, UK\\
$^{5}$School of Physics \& Astronomy, University of Birmingham, Edgbaston, Birmingham, B15 2TT, UK\\
$^{6}$Physics Institute, University of Bern, Gesellsschaftstrasse 6, 3012 Bern, Switzerland\\
$^{7}$ Center for Space and Habitability, University of Bern, Gesellsschaftstrasse 6, 3012 Bern, Switzerland\\
$^{8}$ DTU Space, National Space Institute, Technical University of Denmark, Elektrovej 328, DK-2800 Kgs. Lyngby, Denmark\\
$^{9}$Dipartimento di Fisica e Astronomia “Galileo Galilei,” Universitá di Padova, Vicolo del l’Osservatorio 3, I-35122 Padova, Italy\\
$^{10}$Center for Astrophysics | Harvard \& Smithsonian, 60 Garden Street, Cambridge, MA 02138, USA\\
$^{11}$Département d’astronomie de l’Université de Genève, Chemin Pegasi 51, CH-1290 Versoix, Switzerland\\
$^{12}$Space sciences, Technologies and Astrophysics Research (STAR) Institute, University of Liège, Allée du 6 Août 19C, 4000 Liège, Belgium\\
$^{13}$Observatoire de Genève, University of Geneva, Chemin Pegasi 51, 1290 Versoix, Switzerland\\
$^{14}$Mullard Space Science Laboratory, University College London, Holmbury St Mary, Dorking, Surrey, RH5 6NT, UK\\
$^{15}$INAF - Osservatorio Astronomico di Palermo, Piazza Del Parlamento 1, 90134 Palermo\\
$^{16}$Fundación Galileo Galilei - INAF, Rambla J. A. F. Perez, 7, E-38712 S.C. Tenerife, Spain\\
$^{17}$Astrophysics Group, University of Exeter, Exeter EX4 2QL, UK\\
$^{18}$STFC Ernest Rutherford Fellow\\
$^{19}$Astrophysics Group, Cavendish Laboratory, University of Cambridge, J.J. Thomson Avenue, Cambridge CB3 0HE, UK\\
$^{20}$INAF - Osservatorio Astrofisico di Torino, Via Osservatorio 20, I-10025 Pino Torinese, Italy\\
$^{21}$Department of Physics and Kavli Institute for Astrophysics and Space Research, Massachusetts Institute of Technology, Cambridge, MA 02139, USA
}
\date{Accepted XXX. Received YYY; in original form ZZZ}

\pubyear{2023}

\begin{document}
\label{firstpage}
\pagerange{\pageref{firstpage}--\pageref{lastpage}}
\maketitle

\begin{abstract}
We report on the detailed characterization of the HD 77946 planetary system. HD 77946 is an F5 ($M_*$ = 1.17 M$_{\odot}$, $R_*$ = 1.31 R$_{\odot}$) star, which hosts a transiting planet recently discovered by NASA's Transiting Exoplanet Survey Satellite (\textit{TESS}), classified as TOI-1778 b. Using \textit{TESS} photometry, high-resolution spectroscopic data from HARPS-N, and photometry from \textit{CHEOPS}, we measure the radius and mass from the transit and RV observations, and find that the planet, HD 77946 b, orbits with period $P_{\rm b}$ = $6.527282_{-0.000020}^{+0.000015}$ d, has a mass of $M_{\rm b} = 8.38\pm{1.32}$M$_\oplus$, and a radius of $R_{\rm b} = 2.705_{-0.081}^{+0.086}$R$_\oplus$. From the combination of mass and radius measurements, and the stellar chemical composition, the planet properties suggest that HD 77946 b is a sub-Neptune with a $\sim$1\% H/He atmosphere. However, a degeneracy still exists between water-world and silicate/iron-hydrogen models, and even though interior structure modelling of this planet favours a sub-Neptune with a H/He layer that makes up a significant fraction of its radius, a water-world composition cannot be ruled out, as with $T_{\rm eq} = 1248^{+40}_{-38}~$K, water may be in a supercritical state. The characterisation of HD 77946 b, adding to the small sample of well-characterised sub-Neptunes, is an important step forwards on our journey to understanding planetary formation and evolution pathways. Furthermore, HD 77946 b has one of the highest transmission spectroscopic metrics for small planets orbiting hot stars, thus transmission spectroscopy of this key planet could prove vital for constraining the compositional confusion that currently surrounds small exoplanets.

\end{abstract}




\section{Introduction}

The Transiting Exoplanet Survey Satellite (\textit{\textit{TESS}}) \citep{Ricker2014}, launched in April 2018, has discovered thousands of planetary candidates in the half a decade its been operating and led to many precise planetary radius measurements. The CHaracterising ExOPlanet Satellite (\textit{CHEOPS}) \citep{Benz2021}, launched in December 2019, has allowed us to carry out follow-up observations of many \textit{\textit{TESS}} candidates, using ultra-high precision photometry on the bright stars already known to host planets.
However, to determine these planets' bulk compositions, and thus constrain planetary formation and evolution pathways, precise measurements of their masses are also needed. For small transiting planets, dynamical techniques or high precision spectra from ground-based instruments are commonly used. Examples of such instruments include the High Accuracy Radial Velocity Planet Searcher (HARPS) spectrograph hosted by the European Southern Observatory (ESO) 3.6 m telescope in Chile \citep{Mayor_2003}; and its Northern hemisphere counterpart, HARPS-N, on the Italian 3.58 m Telescopio Nazionale Galileo on La Palma Island \citep{Cosentino_2012}.  Both HARPS and HARPS-N have been shown to have short-term precisions of $\sim$30 cms\textsuperscript{-1} and sub-ms\textsuperscript{-1} precisions over time-scales of months \citep{Consentino_2014}.

The ‘radius valley’ that separates super-Earths and sub-Neptunes has been consistently observed at $\sim$1.5 - 2 R$_{\oplus}$ \citep{Fulton_2017, Van_Eylen_2018, FultonPetigura2018, Ho_2023}, and is largely without planets. Debate currently surrounds the origin of this gap, with proposed scenarios including core-powered mass loss, photoevaporation, or that these planets are primordially rocky. Interpretations differ on the physical mechanism of atmospheric mass-loss, but the result is the same - primordially accreted atmospheres are removed in such a way that different planets are affected in different ways over different timescales, resulting in a valley which separates a population of stripped-core planets (super-Earths) from those which have retained their H/He envelopes (sub-Neptunes) \citep{owen_kepler_2013, Ginzburg2018, Gupta2019}. Whilst investigations have been launched into the existence of planets in the valley, it is still not immediately clear if there are any firm detections of planets in the gap. Larger population studies have indicated that at least a few planets fall within the radius valley, but conversely smaller and more precise planet samples have revealed a complete lack of planets across the gap \citep{Ho_2023}. It has been observed, however, that the valley varies with planetary orbital separation, system age, and mass and metallicity of the host stars, which is consistent with the predictions of atmospheric mass-loss models \citep{David2021, Sandoval2021, Petigura2022}.

The internal structure of sub-Neptunes is not just limited to that of a rocky core surrounded by a gaseous atmosphere, it has been theorised that these planets might hold significant fractions of ices or liquid water \citep{Rogers2010, Dorn2017}. \citet{Zeng2021} suggested that the radii of planets hotter than 900K and with masses below 20 M$_{\oplus}$ can be reproduced assuming ice-dominated compositions without significant gaseous envelopes. The entire mass-radius (M-R) distribution of planets around the radius valley could therefore be separated into three groups: purely rocky planets, gas-poor water worlds, and gas-rich water worlds, where the first two represent the two peaks in the radius distribution \citep{Luque_2022}. However, \citet{Rogers2023} argues that the existence of small planets with hydrogen atmospheres are consistent with the data, once thermal evolution and mass-loss are properly accounted for. This means that there is a strong degeneracy between water-world and silicate/iron-hydrogen models, and that the characterisation of larger sub-Neptunes that reside in this area of the M-R diagram can be used to determine planetary evolution and formation pathways \citep{Kubyshkina2022}.

Here, we present the analysis of the \textit{TESS} target HD 77946 (also known as TIC39699648, HIP 44746). We gathered radial velocity (RV) data from HARPS-N, as well as photometric data from \textit{CHEOPS} and \textit{TESS}, to measure its mass and radius and interpret its composition and investigate whether it meets current radius valley and compositional expectations. Currently, only about $\sim$11\% of sub-Neptunes have a measured mass, due to the difficulty in measuring their small RV signals \citep{Kubyshkina2022}. In this paper we add a significant contribution to this tiny sample of well-characterised small planets.

The paper is organised as follows. In Section \ref{sec:observations}, we detail the \textit{TESS} and \textit{CHEOPS} transit observations, and the HARPS-N RV observations. In Section \ref{sec:stellar}, we derive the parameters of the host star, by combining high resolution spectra with other sources of ancillary information. In Section \ref{sec:modelling}, we describe the approach to modelling the HD 77946 planetary system. In Section \ref{sec:discussion}, we show the resulting properties of HD 77946 b and discuss its composition in comparison to radius valley predictions and current compositional theories. Then to conclude, in Section \ref{sec:conclusion}, we provide a brief summary and final conclusions.

\section{Observations}
\label{sec:observations}
\subsection{Photometry}

\subsubsection{\textit{TESS}}

HD 77946 (also known as TOI-1778) was observed first by \textit{TESS} in sector 21 (year 2) of its primary mission \citep{Ricker2014}, between UTC January 21 2020 and UTC February 18 2020 (BJD 2458870.43 and BJD 2458897.78 respectively). At this point TOI-1778.01 was identified as a \textit{TESS} Object of Interest (TOI), due to a possible transiting planet identified by the Science Processing Operations Center/Transiting Planet Search (SPOC/TPS) pipeline as outlined in \citet{Guerrero2021}. The SPOC pipeline \citep{Jenkins2016} detected a candidate planetary signal with a period of 6.52\,days, which was verified by a classification algorithm (\textit{TESS}-ExoClass (TEC)\footnote{\url{https://github.com/christopherburke/TESS-ExoClass}}) and then vetted by the \textit{TESS} team, thus promoting it to TOI status. The target was then re-observed during the \textit{TESS} extended mission in sector 47 (year 4) between UTC December 30 2021 and UTC January 28 2022 (BJD 2459579.80 and BJD 2459600.95). Over a total of 54.8 days, \textit{TESS} obtained 39,238 images of the target with 2 minute cadence. The data observed in sector 21 were taken on CCD 2 of camera 1, whilst the data observed in sector 47 were taken on CCD 1 of camera 1. The data were reduced by the \textit{TESS} data processing pipeline developed by the SPOC \citep{Jenkins2016}, which identified a total of 7 transits of TOI-1778.01 over the 54.8 day baseline, with a signal-to-noise (SNR) ratio of 15.7 over both sectors.

\begin{figure}
    \centering
    \includegraphics[width=\linewidth]{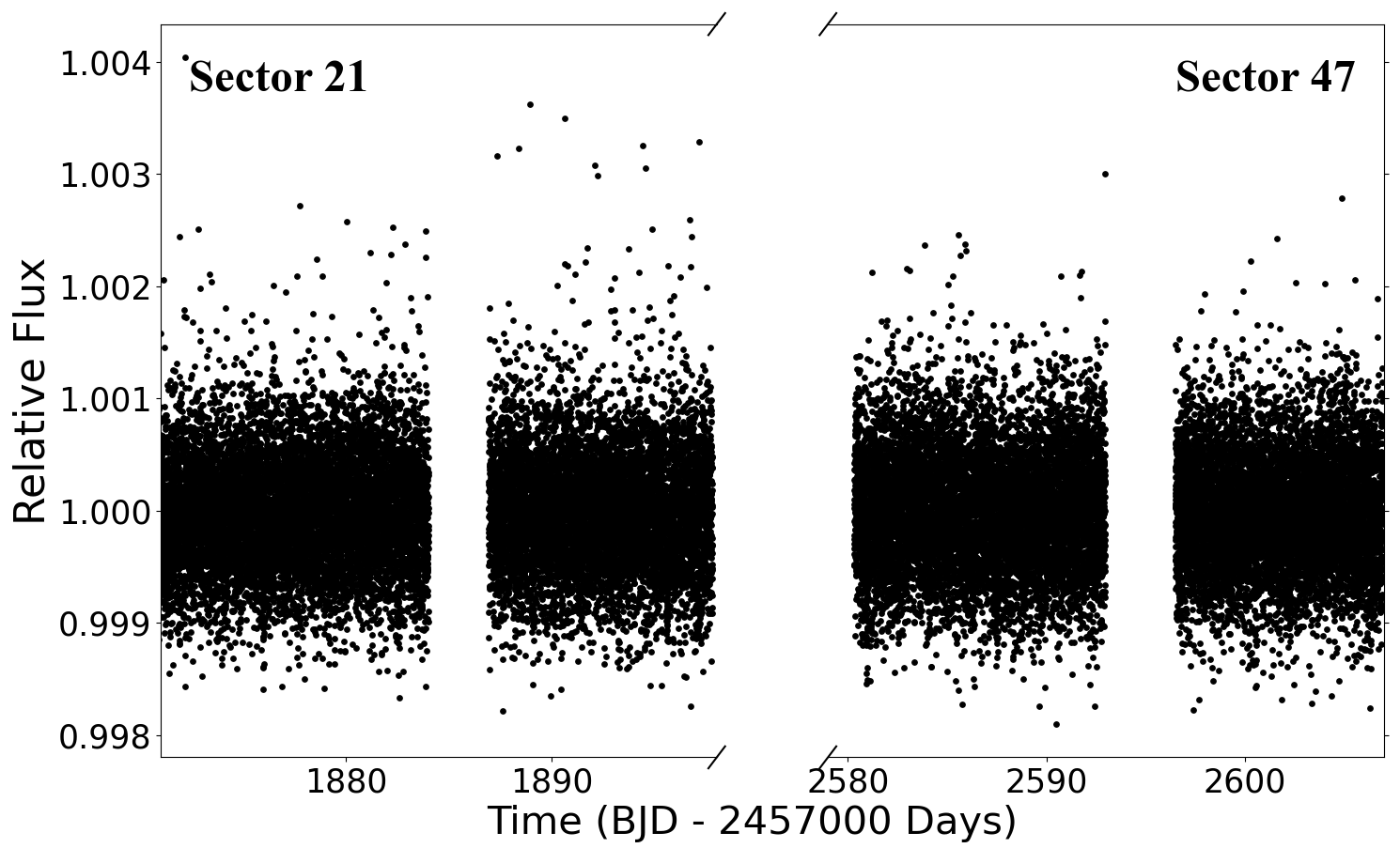}
    \caption{\textit{TESS} photometry of HD 77946 from sectors 21 and 47. Figure shows the PDCSAP \textit{TESS} light curve where systematic errors have been removed, but the resulting light curve has not been corrected for low-frequency variations such as stellar activity.}
    \label{fig:TESS_LC}
\end{figure}

For the photometric analysis, we downloaded the \textit{TESS} photometry from the Mikulski Archive for Space Telescopes (MAST)\footnote{\url{https://mast.stsci.edu/portal/Mashup/Clients/Mast/Portal.html}} and started our analysis using the Pre-search Data Conditioning Simple Aperture Photometry (PDCSAP) light curve reduced by the SPOC. In the initial Simple Aperture Photometry (SAP) flux, a $0.94$\,day gap exists in sector 21, between BJD 2458884.00 and BJD 2458884.94, as well as a $0.97$\,day gap in sector 47, between  BJD 2459592.96 and BJD 2459593.93. Both of these are due to telescope reorientation for transmission of data. This caused spikes in the SAP flux immediately after the telescope reorientation, between BJD 2458884.94 and BJD 2458886.92, and BJD 2459593.93 and BJD 2459596.48 respectively. These gaps and subsequent spikes were removed when deriving the PDCSAP light curve. Hence, the remaining PDCSAP light curve, shown in Figure \ref{fig:TESS_LC}, contains a $2.92$\,day gap between BJD 2458884.00 and BJD 2458886.92, and a $3.52$\,day gap between BJD 2459592.96 and BJD 2459596.48. Throughout the remainder of this work we use the PDCSAP fluxes for determining any stellar and planetary signals.

In Section \ref{sec:fit}, we describe our approach to the modelling of the \textit{TESS} photometry.

\subsubsection{\textit{CHEOPS}}

The \textit{CHEOPS} spacecraft \citep{Benz2021} was launched in 2019 to conduct ultra-high precision photometry to both aid in the discovery of new planets \citep{Leleu2021,Osborn2022,Serrano2022,Wilson2022} and refine their properties \citep{Bonfanti2021,Delrez2021,Lacedelli2022}. To confirm and characterise the planet in the HD 77946 system we obtained 2 visits of a transit of planet b within the \textit{CHEOPS} AO-2 Guest Observers programme ID:07 (PI: Mortier), with the log of these observations presented in Table \ref{tab:cheops_obs}.

\begin{figure}
    \centering
    \includegraphics[width=\linewidth]{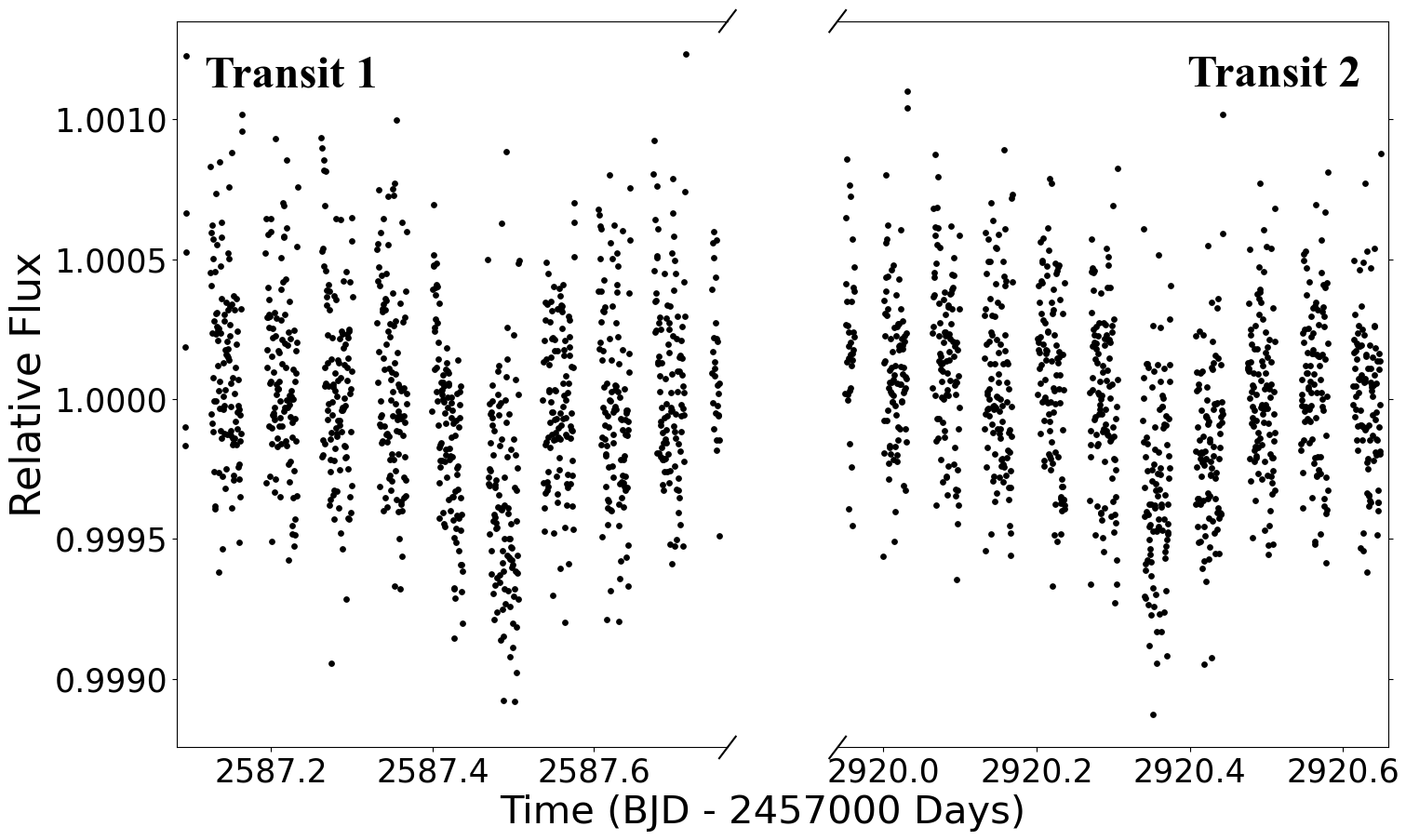}
    \caption{\textit{CHEOPS} photometry of HD 77946 from the 2 transits observed. Figure shows the \textit{CHEOPS} light curves which have been decorrelated with the parameters in Table \ref{tab:cheops_obs}.}
    \label{fig:CHEOPS_LC}
\end{figure}

The data were processed using the \textit{CHEOPS} Data Reduction Pipeline (DRP v13; \citealt{Hoyer2020}) that conducts frame calibration, instrumental and environmental correction, and aperture photometry using pre-defined radii ($R$ = 22.5\arcsec [RINF], 25.0\arcsec [DEFAULT], 30.0\arcsec [RSUP], and OPTIMAL radius) as well as a noise-optimised radius [ROPT]. The size of the OPTIMAL radius is determined on a per-visit basis to account for the differing brightnesses of different targets, and is determined by minimizing the signal-to-noise ratio \citep{Hoyer2020}. Both visits of HD 77946 have the same optimized aperture radius of 25 pixels. The DRP produced flux contamination (see \citealt{Hoyer2020} and \citealt{Wilson2022} for computation and usage) that was subtracted from the light curves. We retrieved the data and corresponding instrumental basis vectors, assessed the quality using the {\sc pycheops} Python package \citep{Maxted2022}, and decorrelated with the parameters suggested by this package, which can be found in Table \ref{tab:cheops_obs}. Outliers were also trimmed from the light curves, with points that were 4$\sigma$ away from the median value removed. We concluded that using the OPTIMAL radius was the best option as it gave the highest signal-to-noise. Therefore, we used these detrended data, shown in Figure \ref{fig:CHEOPS_LC}, for further analysis.

\begin{table*}
    \centering
    \renewcommand\multirowsetup{\raggedright}
    \caption{Log of \textit{CHEOPS} Observations of HD 77946 b.  The column T\textsubscript{exp} gives the exposure time in terms of the integration time per image multiplied by the number of images stacked on-board prior to download. N\textsubscript{obs} is the number of frames. Effic. is the proportion of the time in which unobstructed observations of the target occurred. R\textsubscript{ap} is the aperture radius used for the photometric extraction. RMS is the standard deviation of the residuals from the best fit. The variables in the final column are as follows: time, t; spacecraft roll angle, $\phi$, PSF centroid position, (x, y); smear correction, \texttt{smear}; aperture contamination, \texttt{contam}; image background level, \texttt{bg}.}
    \begin{tabular}{llllllllll}
         \hline\hline 
         \\
         Start Date & Duration & T\textsubscript{exp} & N\textsubscript{obs} & Effic. & File key & APER & R\textsubscript{ap} & RMS & Decorrelation \\ 
         (UTC) & (h) & & & (\%) & & & (pixels) & (ppm) & \\
         \hline 
         \multicolumn{10}{l}{} \\
         2022-01-07T14:06 & 15.89 & 1 x 30s & 1064 & 56 & \multirow{2}{60pt}{CH\_PR220007 \_TG001301\_V0200} & OPTIMAL & 25 & 1656 & sin($\phi$), cos($\phi$), sin(2$\phi$), cos(2$\phi$) \\ \\
         2022-12-06T10:42 & 16.77 & 1 x 30s & 1096 & 54 & \multirow{2}{60pt}{CH\_PR220007 \_TG001401\_V0200} & OPTIMAL & 25 & 785 & t, t$^2$, x, y, sin($\phi$), cos($\phi$), \\ & & & & & & & & & sin(2$\phi$), cos(2$\phi$), sin(3$\phi$), \\ & & & & & & & & & cos(3$\phi$), \texttt{contam, smear, bg}  \\ 
         \hline  
    \end{tabular}
    \label{tab:cheops_obs}
\end{table*}

\subsection{HARPS-N Spectroscopy}

\begin{figure}
    \centering
    \includegraphics[width=\linewidth]{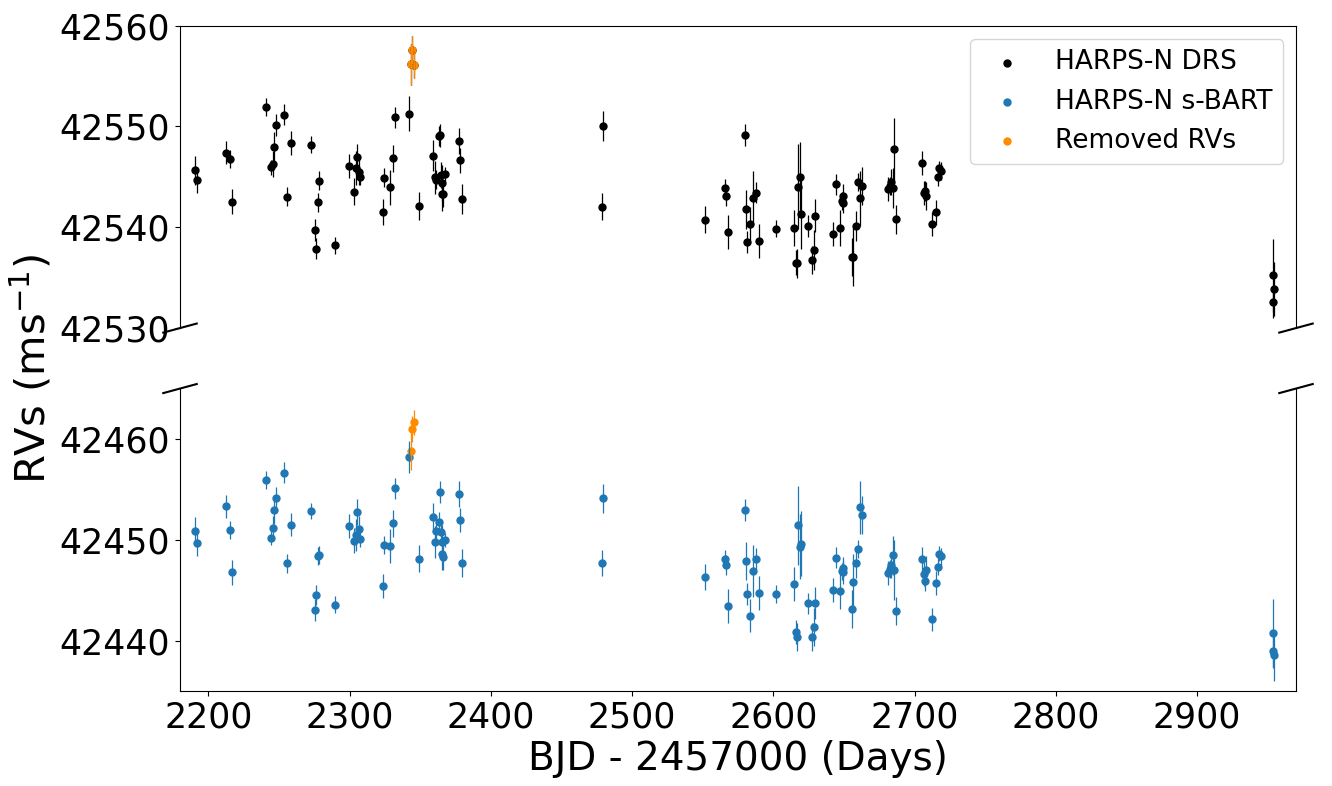}
    \caption{The HARPS-N DRS and s-BART radial velocities plotted against time. The DRS reduced RVs are plotted in black whilst the s-BART reduced RVs are plotted in blue. The same three observations in both datasets, which were removed due to issues with the autoguider, are highlighted in orange.}
    \label{fig:RV_plot}
\end{figure}

We collected 102 spectra for HD 77946 with the HARPS-N high resolution spectrograph ($R=115000$) \citep{Cosentino_2012} installed on the 3.6m Telescopio Nazionale Galileo (TNG) at the Observatorio de los Muchachos in La Palma, Spain. HARPS-N observed HD 77946 between UTC December 2020 and UTC January 2023 (BJD 2459190.74 and BJD 2459954.68) as part of the HARPS-N Collaboration’s Guaranteed Time Observations (GTO) programme. Our observational strategy initially consisted of taking one observation per night, lasting 15 minutes, for several consecutive nights in order to sample the RV curve of the transiting planet, and at later times performed more sporadic observations where we aimed to sample less-well resolved sections of the planet's orbital phase. The sampling strategy is shown in Figure \ref{fig:RV_plot}, along with the two different methods used to reduce the spectra and their resultant RVs, as explained in Sections \ref{sec:DRS} and \ref{sec:sBART}. HD 77946 is a bright star, with V-band magnitude V$=8.99\pm0.03$, hence exposure times of $900$s per observation were sufficient.

The obtained spectra had signal-to-noise ratios in the range 33-152 (average SNR = 91). We eliminated three observations, taken on consecutive nights BJD=2459343.44, 2459344.46, 2459345.46, due to an issue with the guiding component of HARPS-N which resulted in an anomalous offset and was confirmed in our well-observed standard stars. The removal of these RVs slightly increased the signal-to-noise (average SNR = 92), but did not change the average RV precision. Additionally, in UTC October 2021 the cryostat on the instrument was replaced leading to an offset between the two seasons, which was accounted for when modelling the data.

\subsubsection{DRS}
\label{sec:DRS}

The spectra were reduced using version 2.3.5 of the HARPS-N Data Reduction Software (DRS) \citet{Dumusque_2021}, with a G2 mask used in the cross-correlation function (CCF). The spectra obtained resulted in an average RV precision of 1.4 ms\textsuperscript{-1} and an RMS of 4.44 ms\textsuperscript{-1}. The HARPS-N DRS data are presented in Table \ref{tab:RV_HARPSN_1} which includes the radial velocities as well as the activity indicators: full width at half maximum (FWHM) of the CCF, the line Bisector Inverse Slope (BIS), and the log R$'_{\rm HK}$ converted from the S-index following \citet{Noyes1984a}.

\subsubsection{s-BART}
\label{sec:sBART}

Following the confident detection of the transiting planet with the DRS RVs, the s-BART \citep{Silva_2022} template-matching pipeline was applied to the same data-set. This approach is built around the core assumption that the RV signal introduced by the presence of an orbiting planet should be achromatic, i.e. independent of the wavelength at which it is measured. Telluric features are accounted for through a binary mask that rejects any wavelength for which Earth's transmittance drops under 1\% of its continuum value, which is enlarged to consider the BERV motion. Furthermore, to ensure that each observation is informed by the same wavelength region, s-BART only uses the spectral orders that are common to all observations in the dataset. This model was implemented in a Bayesian framework, allowing for a straightforward and consistent characterization of the uncertainties associated with the RV extraction, and has been yielding good results for both ESPRESSO data (e.g. \citealt{Faria_2022}) and HARPS-N \citep{Wilson2024}. This resulted in an average RV precision of 1.4 ms\textsuperscript{-1} and an RMS of 4.45 ms\textsuperscript{-1}. The HARPS-N s-BART data are presented in Table \ref{tab:RV_HARPSN_3}.

\section{Stellar parameters}
\label{sec:stellar}

\subsection{Atmospheric parameters}

Two independent methods were employed to determine the stellar atmospheric parameters using the high resolution HARPS-N spectra, namely ARES+MOOG and Stellar Parameter Classification (SPC).

The ARES+MOOG method, described in detail in \cite{sousa2014}, determines the stellar parameters using the equivalent widths (EW) of a series of iron absorption lines. For this analysis, all of the HARPS-N spectra were stacked together to create a very high-SNR spectrum. The EWs of a list of neutral and ionised iron absorption lines \citep{sousa2011} are automatically measured using the \texttt{ARESv2} code \citep{sousa2015}. The EWs are then fed into the 2017 version of the radiative transfer code \texttt{MOOG} \citep{sneden1973} where iron abundances are compared to a grid of Kurucz Atlas 9 plane parallel model atmospheres \citep{kurucz1993}, which assume local thermodynamic equilibrium (LTE). The appropriate atmospheric model is determined through a minimisation process \citep{press1992}, from which the best-fit stellar parameters are obtained. Derived values for $\mathrm{log(}g\mathrm{)}$ were subsequently corrected following the method described in \cite{mortieretal14}, and our errors were inflated to account for statistical tension by adding our precision errors and systematic errors in quadrature. We determined values $T_{\rm eff}=6154\pm81$\,K, $\mathrm{log(}g\mathrm{)} = 4.41\pm0.11$, $\mathrm{[Fe/H]}=0.24\pm0.06$. Using these atmospheric parameters, we further used ARES+MOOG to derive individual elemental abundances for magnesium and silicon. These elemental abundances, in combination with the iron abundance, can be used to derive the planet's interior composition. Details of the method and line list are found in \citet{Mortier2013}. We ran \texttt{MOOG} again in LTE mode and the values of \citet{Asplund2009} were used as a reference for the solar values.

By comparison SPC, described in detail in \cite{buchhave2012,buchhave2014}, is a spectral synthesis method. In this method, each HARPS-N spectrum is analysed individually and final stellar parameters are calculated as the average of each individual result weighted by their signal to noise. The value for $\mathrm{log(}g\mathrm{)}$ is additionally constrained by using isochrones. Again, our precision errors and systematic errors are added in quadrature. We obtained values $T_{\rm eff}=5937\pm49$\,K, $\mathrm{log(}g\mathrm{)} = 4.20\pm0.10$, $\mathrm{[m/H]}=0.12\pm0.08$. We note that the iron abundance derived by this method is similar to an overall metallicity due to the abundance of iron lines in an optical spectrum.

Our final parameters adopted here, displayed in Table \ref{tab:stellarparameters}, are the average of the two methods. We used a Monte Carlo approach, using 100000 samples to generate random Gaussian distributions of the three relevant parameters. An average distribution was then computed for each parameter from which the mean and standard deviation are taken as the final adopted atmospheric parameters.

We calculate the U, V, and W Galactic velocities using the method of \cite{Johnson1987} and data from the most recent \textit{Gaia} data release (for examples see \citet{mortieretal2020,Wilson2022}), and compute the thin and thick disk, and halo membership probabilities. This was done using the framework outlined in \citet{Wilson2024}, which combines new \textit{Gaia} offset-corrected \citep{Lindegren2021} data \citep{GaiaDR3} with a known probability method \citep{Reddy2006}, then uses a novel Monte Carlo approach to compute a weighted average combination of probabilities using four sets of velocity standards \citep{Bensby2003, Bensby2014,Reddy2006,Chen2021} corrected for the Local Standard of Rest \citep{Koval2009}. We find U $= 44.66 \pm 0.27$ kms\textsuperscript{-1}, V $= -2.36 \pm 0.05$ kms\textsuperscript{-1},
W $= 12.88 \pm 0.25$ kms\textsuperscript{-1}, resulting in 98.5\%, 1.5\%, and 0\% thin disk, thick disk, and halo probabilities, respectively, which is in line with the given metallicity for this type of star.

\begin{table}
    \centering
    \caption{Stellar parameters for HD 77946.}
    \renewcommand\multirowsetup{\raggedright}
    \begin{tabular}{lll}
         \hline\hline 
         \\
         Parameter & Value & Source \\
         \hline
         {\textit{Name}} \\
         2MASS & $J09070675+4640214$ & \citetalias{2MASS}\\
         \textit{Gaia} DR3 & $1011435012611767552$ & \citetalias{GaiaDR3} \\
         HD & $77946$ & \citet{Nesterov1995}\\
         HIP & $44746$ & \citet{Perryman1997}\\
         TIC & $39699648$ & \citet{Stassun2018}\\
         TOI & $1778$ & \citet{Guerrero2021}\\
         TYC & $3424-01533-1$ & \citetalias{Tycho2}\\
         \textit{WISE} & $J090706.66+464020.9$ & \citetalias{Wright2010}\\
         \hline
         \multicolumn{3}{l}{\textit{Photometry}} \\
         RA & 09:07:06.63 & \citetalias{GaiaDR3}\\
         Declination & +46:40:20.73 & \citetalias{GaiaDR3}\\
         $\pi$\,[mas] & $10.07\pm0.02$ & \citetalias{GaiaDR3} \\
         $d$\,[pc] & $99.0^{+0.3}_{-0.2}$ & \citet{Bailer2021}\\
         log $R'_{\rm HK}$ & $-5.01 \pm 0.05$ & This work \\
         P$_{\rm rot}$ [d] & $14\pm2$ & This work\\
         RV [m s$^{-1}$] & $2.22_{-0.21}^{+0.06}$ & This work\\
         $U$ [km s$^{-1}$] & $+44.66\pm0.27$ & This work \\
         $V$ [km s$^{-1}$] & $-2.36\pm0.05$ & This work\\
         $W$ [km s$^{-1}$] & $+12.88\pm0.25$ & This work \\
         \hline
         Spectral Type & F5 & \citet{Cannon1993}\\
         \hline
         $B$ [mag] & $9.59\pm0.02$ & \citetalias{Tycho2}\\
         $V$ [mag] & $9.00\pm0.02$ & \citetalias{Tycho2}\\
         $J$ [mag] & $7.904\pm0.02$ & \citetalias{2MASS} \\
         $H$ [mag] & $7.67\pm0.02$ & \citetalias{2MASS} \\
         $K$ [mag] & $7.62\pm0.02$ & \citetalias{2MASS} \\
         $W1$ [mag] & $7.56\pm0.03$ & \citetalias{Wright2010} \\
         $W2$ [mag] & $7.64\pm0.02$ & \citetalias{Wright2010} \\
         $W3$ [mag] & $7.62\pm0.02$ & \citetalias{Wright2010} \\
         \hline
         \multicolumn{3}{l}{\textit{Atmospheric parameters}} \\
         $T_{\rm eff}$\,[K] & $6154\pm81$ & ARES+MOOG \\
         $\mathrm{log} g$\,[cgs] & $4.41\pm0.11$ & ARES+MOOG \\
         $[\mathrm{Fe/H}]$ & $0.24\pm0.06$ & ARES+MOOG \\
         $[\mathrm{Mg/H}]$ & $0.41\pm0.22$ & ARES+MOOG \\
         $[\mathrm{Si/H}]$ & $0.31\pm0.08$ & ARES+MOOG \\
         $\xi_t$\,[kms\textsuperscript{-1}] & $1.42\pm0.07$ & ARES+MOOG \\
         $T_{\rm eff}$\,[K] & $5937\pm49$ & SPC \\
         $\mathrm{log} g$\,[cgs] & $4.20\pm0.10$ & SPC \\
         $[\mathrm{m/H}]$ & $0.12\pm0.08$ & SPC \\
         $v \mathrm{sin} i$\,[kms\textsuperscript{-1}] & $3.5\pm0.5$ & SPC \\
         \multicolumn{3}{l}{\textit{Adopted average parameters}} \\
         $T_{\rm eff}$\,[K] & $6046\pm50$ & ARES+MOOG/SPC \\
         $\mathrm{log} g$\,[cgs] & $4.30\pm0.07$ & ARES+MOOG/SPC \\
         $[\mathrm{m/H}]$ & $0.18\pm0.05$ & ARES+MOOG/SPC \\
         \hline
         \multicolumn{3}{l}{\textit{Stellar mass, radius, and age}} \\
         $M_*$\,[M$_{\odot}$] & $1.17^{+0.06}_{-0.05}$ & \texttt{isochrones} \\[0.15cm]
         $R_*$\,[R$_{\odot}$] &  $1.31\pm0.01$ & \texttt{isochrones} \\
         $t_\ast$\,[Gyr] & $3.85^{+1.26}_{-1.61}$ & \texttt{isochrones} \\[0.15cm]
         $\mathrm{log} g$\,[cgs] & $4.27^{+0.03}_{-0.02}$ & \texttt{isochrones}\\
         \hline
         
    \end{tabular}
    \label{tab:stellarparameters}
\end{table}

\subsection{Stellar mass, radius, and age}\label{sec:stellarmassradius}

Stellar parameters of HD 77946 were determined using \texttt{isochrones} and stellar evolution tracks. For a detailed discussion on this see \citet{mortieretal2020}. In brief, spectroscopically determined values for the effective temperature and metallicity from ARES+MOOG and SPC were used as inputs to the \texttt{isochrones} python package \citep{morton2015}, together with the \textit{Gaia} Early Data Release 3 \citep{Brown_2021} parallax and the apparent magnitude in 8 bands, listed in Table \ref{tab:stellarparameters}. Two sets of stellar evolution models were used, the Dartmouth Stellar Evolution Database \citep{dotteretal08} and the MESA Isochrones and Stellar Tracks \citep[MIST - ][]{dotteretal16}. For each stellar evolution model (Dartmouth/MIST) the analysis was run twice; once using each of the spectroscopic inputs (ARES+MOOG/SPC), totalling to 4 individual runs. Final values for the stellar mass, radius, and age were obtained by combining the posteriors from each run, and taking the median and the 16th and 84th percentile. We derived values $M_{*}=1.17^{+0.06}_{-0.05}$\,M$_{\odot}$, $R_{*}=1.31\pm0.01$\,R$_{\odot}$, and $t_\ast=3.85^{+1.26}_{-1.61}$\,Gyr. Combining the derived mass and radius, we can compute a new value for the surface gravity, which is consistent with, but more precise than the spectroscopic surface gravity.

\subsection{Stellar activity and rotation}
\label{sec:stellaractivity}

Signals intrinsic to stars themselves still pose significant problems, as they can induce RV variations that can drown out or even mimic planetary signals \citep{Rajpaul2021}. Modelling and thus mitigating these signals is notoriously difficult. However, a technique that has proved successful for disentangling stellar activity signals from planetary signals is to use a Gaussian Process (GP) framework \citep{haywood_planets_2014, Rajpaul2015, Stock2020}. GPs have a number of features that make them very well suited to the joint modelling of activity processes and dynamical (e.g. planetary) signals, giving probabilistic estimates of the model parameters, which leads to more precise and accurate mass estimates.

\begin{figure}
    \centering
    \includegraphics[width=\linewidth]{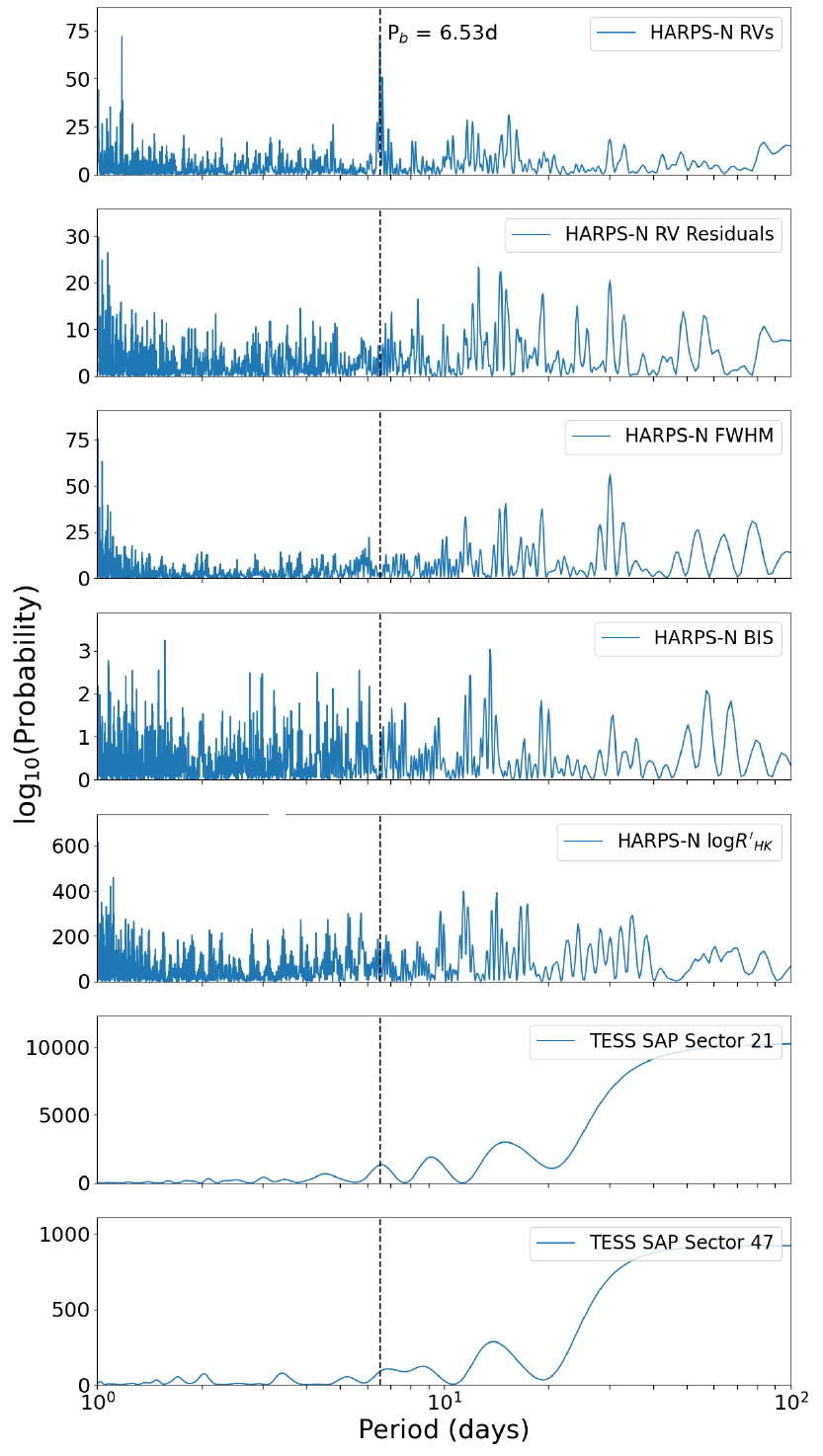}
    \caption{BGLS periodograms calculated using the HARPS-N RVs (first), the RV residuals (second), the FWHMs of the CCF (third), the BIS (fourth), the S$_{\rm HK}$ index (fifth), and the SAP from the two \textit{TESS} sectors (sixth \& seventh). The labelled dashed line shows the period of the transiting planet at 6.53d.}
    \label{fig:BGLS}
\end{figure}

In addition to providing us with RVs, the HARPS-N spectra can be used to infer information about stellar activity. If it is observed that there is a significant stellar activity signal present in the RVs, then including a GP model in the RV fit is advised. We performed a Bayesian generalised Lomb-Scargle (BGLS) \citep{Boisse2011, mortieretal15} analysis on the HARPS-N RVs, RV residuals (where the signal of the transiting planet has been subtracted), and each of the activity parameters, as well as the \textit{TESS} SAP, shown in Figure \ref{fig:BGLS}, in order to ascertain whether this was the case.

The signal of the transiting planet ($P_{\rm b}$) at 6.53 days is clearly visible from the RV values from HARPS-N (first panel of Figure \ref{fig:BGLS}, highlighted by the labelled black dashed line) and \textit{TESS} SAP (sixth and seventh panels of Figure \ref{fig:BGLS}), with no corresponding signal in the RV residuals, FWHM, BIS, or log R$'_{\rm HK}$ values. Whilst the absence of signals in activity indicators does not necessarily imply that the signal is not of stellar origin \citep{CollierCameron2019}, this signal can be verified to be that of the planet as it is consistent with the $P_{\rm b}$ derived from both the \textit{TESS} and \textit{CHEOPS} photometry. Whilst a definitive stellar activity signal was not identified in the stellar activity indicators presented in Figure \ref{fig:BGLS}, there is a forest of peaks in the 10 - 20 day period range which may be indicative of stellar activity. The activity indicators can be sensitive to different types of variability, however, so peaks at the same period in multiple panels are not necessarily correlated. Hence it was deemed to be necessary to include a quasi-periodic GP regression model in the RV fit. A GP model was not used on the \textit{TESS} data as the light curves are too short and any long term trends greater than length of the \textit{TESS} sector would then be excluded.

The S-index and associated log $R'_{\rm HK}$ are traditionally seen as an excellent indicator for a star’s magnetic cycle. The Sun, for example, has an S-index varying between 0.16 and 0.18 throughout its 11-yr magnetic cycle \citep{Egeland2017}. The average log $R'_{\rm HK} = -5.01 \pm 0.05$ for HD 77946, which would indicate a rotation period of $\sim 18$ d using the calibrations from \citet{Noyes1984b}, and a similar rotation period of $\sim 19$ d and age of $3.5$~Gyr using the calibrations from \citet{Mamjek2008}. However, these calibration relations are just estimates and a manifestation of the rotation period in the data may not be the same as the physical rotation period \citep{Nava2020}. The average value of log R$'_{\rm HK}$ can additionally be used to estimate the expected stellar-induced RV variations. From equation (1) of \citet{Hojjatpanah2020}, a value of -5.01 for log $R'_{\rm HK}$ translates to an RV rms of $2.22_{-0.21}^{+0.06}$ ms\textsuperscript{-1}. In contrast, \citet{Suarez2017} estimates the RV semi-amplitude induced by stellar activity variations of an F star with average log $R'_{\rm HK}$ of -5.01 should be lower than 1 ms\textsuperscript{-1}, though this was based on a smaller sample. HARPS-N solar data shows an rms of 1.63 ms\textsuperscript{-1} \citep{CollierCameron2019} whilst the Sun was approaching solar minimum with values of log $R'_{\rm HK}$ around -4.97, thus we can expect RV variations from stellar activity at the level of 0.5-3 ms\textsuperscript{-1}.

There is a risk that when a model is used to try and constrain the planetary mass, some of the stellar activity is also modelled, resulting in a mass much higher or lower than the true value \citep{Nava_2022}. \cite{CollierCameron_2021} developed an algorithm called {\sc scalpels} to combat this problem, by separating the stellar RV variability component driven by spectral line-shape changes while preserving the planetary shift signals \citep{Wilson2022}. We used the {\sc tweaks} pipeline, described in \citet{AncyAnna2022, AnnaJohn2023}, for performing the  stellar activity mitigation in the wavelength and time domain simultaneously. This is achieved by incorporating the {\sc scalpels} basis vectors representing the shape components in the nested sampling package called {\sc kima}\footnote{\url{https://github.com/j-faria/kima}} \citep{Kima2018} for stellar activity decorrelation. As \citet{AncyAnna2022} have found, some shift-like signals elude {\sc scalpels} analysis, so any remaining rotationally-modulated signals were modelled with GP regression applied to the RVs.

We used a quasi-periodic GP kernel, which has four hyperparameters ($\theta, h, \omega , \lambda$) and is described by

\begin{equation}
    \Sigma_{i,j} = h^2\exp\bigg\{-\frac{\sin ^2 [\pi(t_i - t_j)/\theta]}{2\omega^2} - \left(\frac{t_i - t_j}{\lambda}\right)^2\bigg\}
    \label{quasi_periodic}
\end{equation}

where $t_i$ and $t_j$ are two times of observation, and the other terms are the variability period of the star ($\theta = P_{\rm rot}$), the variability amplitude ($h = H_{\rm amp}$), the periodic characteristic length (associated with the number of spots/spot regions on the surface of the star, $\omega = O_{\rm amp}$), and the non-periodic characteristic length (associated with the spot decay timescale, $\lambda$). The GP was run with log-uniform priors on the amplitude 0 $<$ ln($h$) $<$ 2.7 and active-region evolution timescale 24 $<\lambda<$ 40\,d. These boundaries were selected based on the stellar type of HD 77946, an F dwarf, so we assume its active regions evolve on approximately the same timescale as the Sun. The SPC gives $v \sin i = 3.5 \pm 0.5$ kms$^{-1}$, and based on a stellar radius of 1.31 R$_\odot$ this limits the stellar rotation period to a value of below 22 days. Based on this estimate and a few initial runs with uninformative priors, the GP priors on the stellar rotation period were set to be uniform with 12 $<\theta<$ 19\,d. The log harmonic complexity was controlled with a uniform prior -1 $<$ ln($\omega$) $<$ 0. In short, we used a model with up to five unknown Keplerian signals to maximise exploration, and stellar activity decorrelation performed against the {\sc scalpels} basis vectors, incorporated with a quasi-periodic GP regression.

The joint posteriors showed clear detection of a Keplerian signal at orbital period $\sim$ 6.52 days, as well as identified a stellar rotation of $P_{\rm rot} = 13.48_{-0.42}^{+0.98}$ days. From this investigation with the {\sc scalpels} algorithm, we were able to place a prior on $P_{\rm rot}$ of $\mathcal{N}(14.0, 2.0)$ with a boundary of 10-20 days (determined from the activity signals in Figure \ref{fig:BGLS}) for the joint fit with the photometric data, radial velocity data, and the GP. This means that the stellar activity could be properly accounted for and the planetary mass correctly constrained. 

\section{Planet Modelling}
\label{sec:modelling}

\subsection{Joint Transit \& RV Modelling}
\label{sec:fit}

\begin{table*}
    \centering
    \caption{Stellar and planetary properties calculated from the combined RV, transit and GP fit. The prior label $\mathcal{N}$ represents a Normal distribution.}
    \begin{tabular}{lllll}
         \hline\hline 
         \\
         Parameter & Description & Priors & HD 77946 b & HD 77946 b \\  &  &  & (DRS RVs) & (s-BART RVs) \\
         \hline
         \multicolumn{4}{l}{\textit{Stellar parameters}} \\
         \vspace{1mm}
         $\rho_*$\,[g/cm$^3$] & Stellar density & $\mathcal{N}(0.52,0.1)$ & $0.558_{-0.071}^{+0.072}$ & $0.558_{-0.071}^{+0.072}$ \\
         \vspace{1mm}
         $C1$ [TESS] & LD coefficient & $\mathcal{N}(0.27,0.1)$ & $0.32_{-0.10}^{+0.11}$ & $0.35_{-0.10}^{+0.10}$\\
         \vspace{1mm}
         $C2$ [TESS] & LD coefficient & $\mathcal{N}(0.30,0.1)$ & $0.34_{-0.10}^{+0.10}$ & $0.36_{-0.10}^{+0.10}$\\
         \vspace{1mm}
         $C1$ [CHEOPS] & LD coefficient & $\mathcal{N}(0.38,0.1)$ & $0.37_{-0.10}^{+0.10}$ & $0.33_{-0.09}^{+0.09}$\\
         \vspace{1mm}
         $C2$ [CHEOPS] & LD coefficient & $\mathcal{N}(0.27,0.1)$ & $0.25_{-0.09}^{+0.10}$ & $0.23_{-0.10}^{+0.09}$\\
         \vspace{1mm}
         $s_j$ & RV jitter & & $1.09_{-0.31}^{+0.30}$ & $0.91_{-0.33}^{+0.30}$\\
         \vspace{1mm}
         $R_0$ & RV offset 1 & & $42546.17_{-0.97}^{+0.98}$ & $42451.22_{-0.97}^{+0.98}$\\
         \vspace{1mm}
         $R_1$ & RV offset 2 & & $42541.29_{-1.08}^{+0.99}$ & $42445.60_{-1.05}^{+1.00}$\\
         \hline
         \multicolumn{4}{l}{\textit{Stellar activity parameters}} \\
         \vspace{1mm}
         $P_{\rm rot}$\,[days] & Rotational period & $\mathcal{N}(14.0,2.0)$ & $13.37_{-0.25}^{+0.34}$ & $13.56_{-0.28}^{+0.26}$\\
         \vspace{1mm}
         $\lambda$\,[days] & Decay time
         & & $29.49_{-0.33}^{+0.35}$ & $29.48_{-0.33}^{+0.35}$\\
         \vspace{1mm}
         $O_{\rm amp}$ & Coherence scale & & $0.495_{-0.003}^{+0.004}$ & $0.495_{-0.003}^{+0.003}$\\
         \vspace{1mm}
         $H_{\rm amp}$ & Variability amplitude & & $2.54_{-0.42}^{+0.49}$ & $2.56_{-0.42}^{+0.49}$\\
         \hline
         \multicolumn{4}{l}{\textit{Orbital parameters}} \\
         \vspace{1mm}
         $P$\,[days] & Orbital period & & $6.527282_{-0.000021}^{+0.000016}$ & $6.527282_{-0.000020}^{+0.000015}$ \\
         \vspace{1mm}
         $T_{\rm C}$\,[BJD-2450000] & Transit centre & & $9587.4743_{-0.0009}^{+0.0014}$ & $9587.4740_{-0.0008}^{+0.0013}$\\
         \vspace{1mm}
         $T_{14}$\,[hrs] & Transit duration & & $3.289_{-0.490}^{+0.341}$ & $3.551_ {-0.324}^{+0.272}$\\
         \vspace{1mm}
         $R_{\rm b}/R_*$ & Radius ratio & & $0.0194_{-0.0011}^{+0.0012}$ & $0.0189_{-0.0010}^{+0.0011}$\\
         \vspace{1mm}
         $b$ & Impact parameter & & $0.577_{-0.242}^ {+0.133}$ & $0.373_{- 0.231}^{+0.193}$\\
         \vspace{1mm}
         $e$ & Eccentricity & & $0.140_{-0.089}^{+0.092}$ & $0.233_{-0.079}^{+0.064}$\\
         \vspace{1mm}
         $i$\,[deg] & Inclination & & $87.02_{-0.75}^{+1.04}$ & $87.79_{-1.05}^{+1.31}$\\
         \vspace{1mm}
         $a/R_*$ & Semi-major axis ratio & & $11.902_{-0.890}^{+0.730}$ & $12.009_{-0.741}^{+0.650}$\\
         \vspace{1mm}
         $a$\,[AU] & Semi-major axis & & $0.0720_{-0.0012}^{+0.0012}$ & $0.0720_{-0.0012}^{+0.0012}$\\
         \vspace{1mm}
         $K$\,[ms\textsuperscript{-1}] & RV semi-amplitude & & $2.65_{-0.45}^{+0.44}$ & $2.66_{-0.41}^{+0.40}$\\
         \hline
         \multicolumn{4}{l}{\textit{Mass, radius and density}} \\
         \vspace{1mm}
         $M_{\rm b}$\,[M$_{\oplus}$] & Planet mass & & $8.48_{-1.49}^{+1.46}$ & $8.38_{-1.32}^{+1.32}$\\
         \vspace{1mm}
         $R_{\rm b}$\,[R$_{\oplus}$] & Planet radius & & $2.762_{-
         0.097}^{+0.106}$ & $2.705_{-0.081}^{+0.086}$\\
         \vspace{1mm}
         $\rho_{\rm b}$\,[$\rho_{\oplus}$] & Planet density & & $0.426_{-0.086}^{+0.090}$ & $0.450_{-0.081}^{+0.085}$\\

    \end{tabular}
    \label{tab:planetresults}
\end{table*}

To calculate the best-fit planetary and stellar activity values for HD 77946 we performed a combined GP, transit, and RV fit to the \textit{TESS} PDCSCAP, the previously decorrelated \textit{CHEOPS} photometry, and the HARPS-N RVs (with the process separately repeated for both the DRS and s-BART data). We used the \texttt{PyORBIT}\footnote{\url{https://github.com/LucaMalavolta/PyORBIT}, version 8.} code \citep{malavoltaetal16, Malavolta2018}, a package for modelling planetary and stellar activity signals, where best-fit activity and planet parameters were obtained using the dynamic nested sampling code \texttt{dynesty} \citep{speagle2020}. \texttt{PyORBIT} makes use of the \texttt{BATMAN} python package \citep{kreidberg2015} for fitting the transit to the photometric data, where we assumed a quadratic stellar intensity profile for fitting the limb darkening coefficients, and an exposure time of 120.0\,s for the \textit{TESS} observations, and 30.0\,s for the \textit{CHEOPS} observations as inputs to the light curve model. The GP analysis was carried out using a quasi-periodic covariance kernel function as defined in equation \ref{quasi_periodic}, which combines a squared exponential and standard periodic kernel \citep{haywood_planets_2014, Grunblatt2015}. \texttt{PyORBIT} implements this GP quasi-periodic kernel through the \texttt{GEORGE}\footnote{\url{https://github.com/dfm/george}} package \citep{Ambikasaran2016}.
The stellar activity parameters obtained in Section \ref{sec:stellaractivity} were used to inform priors for the GP fit which was performed simultaneously with the combined transit and RV fit.

Inferred stellar and planetary parameters from the combined fit are shown in Table \ref{tab:planetresults}, alongside the priors implemented. The prior for $P_{\rm rot}$ is explained in Section \ref{sec:stellaractivity}, whilst the prior on stellar density comes from stellar mass and radius estimates in Section \ref{sec:stellarmassradius}. The priors placed on the limb-darkening coefficients were calculated using interpolation of Table 25 from \citet{Claret2017} for the \textit{TESS} light curves and Table 8 from \citet{Claret2021} for the \textit{CHEOPS} light curves. The fit converges to a moderately eccentric planetary orbit, with a posterior on $K_{\rm b}$ corresponding to a $\sim6\sigma$ detection. Using the estimates for the stellar mass and radius obtained in Section \ref{sec:stellarmassradius} and the HARPS-N DRS RVs, we derive that HD 77946 b has a mass $M_{\rm b}=8.48_{-1.49}^{+1.46}$\,M$_{\oplus}$ and a radius $R_{\rm b}=2.76_{-0.10}^{+0.11}$\,R$_{\oplus}$. 

We also performed the same combined GP, transit, and RV fit using the s-BART RVs, as mentioned in Section \ref{sec:sBART}, with the resulting inferred stellar and planetary parameters shown in Table \ref{tab:planetresults}. In this case the fit converged to a slightly more eccentric planetary orbit than that of the fit with the DRS RVs, which led to a slightly higher posterior on $K_{\rm b}$ corresponding to a $\sim6.6\sigma$ detection. From this we also derived a slightly lower mass and radius for HD 77946 b of $M_{\rm b}=8.38_{-1.32}^{+1.32}$\,M$_{\oplus}$ and $R_{\rm b}=2.71_{-0.08}^{+0.09}$\,R$_{\oplus}$. These results still fall within $1\sigma$ of those derived from the HARPS-N DRS RVs, but this does suggest that the choice of data reduction technique can influence the resulting planetary parameters. We favour these results as we find that the fit with the s-Bart RVs gives a stronger detection of the planet's mass and required less jitter in the fit, although there is tension in the derived eccentricity when using either RV dataset.

\begin{figure*}
    \centering
    \begin{minipage}[b]{.5\textwidth}
        \centering
        \includegraphics[width=.9\linewidth]{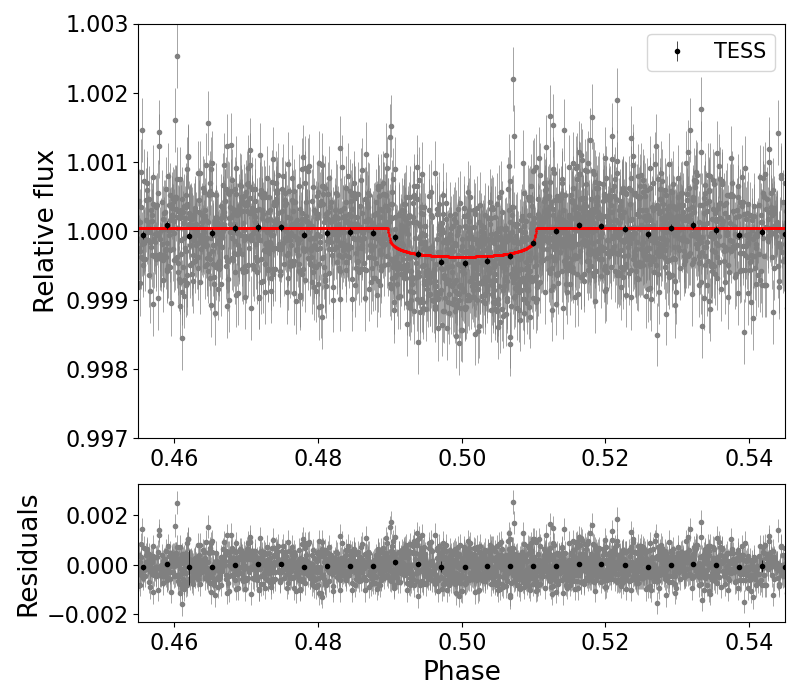}
    \end{minipage}%
    \hfill
    \begin{minipage}[b]{.5\textwidth}
        \centering
        \includegraphics[width=.9\linewidth]{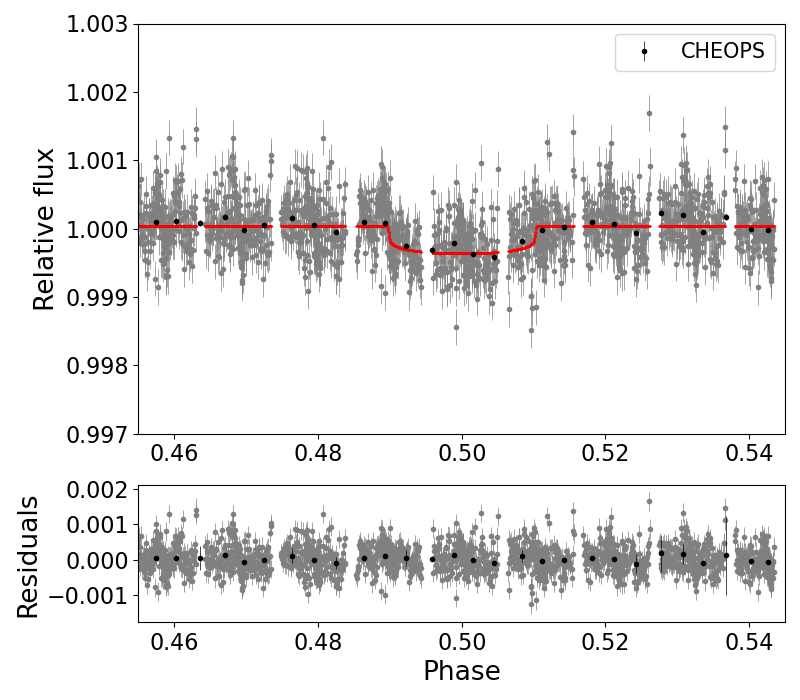}
    \end{minipage}    
    \caption{Combined fit results obtained from \texttt{PyORBIT}. Transit fit to the light curve data from \textit{TESS} (\textit{left}) and \textit{CHEOPS} (\textit{right}). The short cadence (2 minute for \textit{TESS}, 30 second for \textit{CHEOPS}) fluxes are plotted in grey, the fluxes binned every 30 minutes are overplotted in black, and the fitted transit is shown by the red solid line.}
    \label{fig:transit_fit}
\end{figure*}

\begin{figure*}
    \centering
    \begin{minipage}[b]{.5\textwidth}
        \centering
        \includegraphics[width=.95\linewidth]{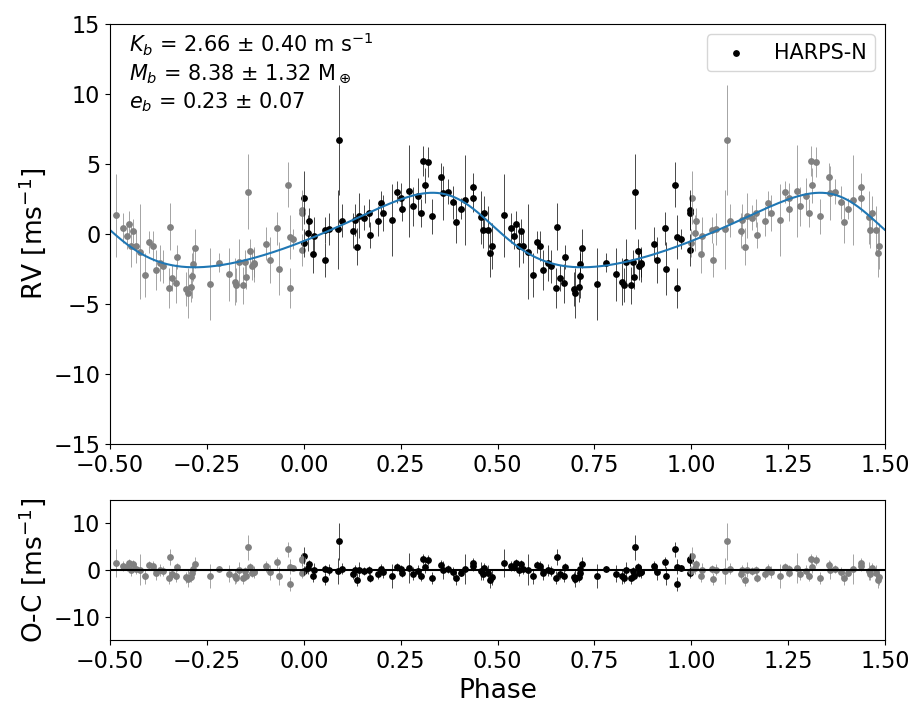}
    \end{minipage}%
    \hfill
    \begin{minipage}[b]{.5\textwidth}
        \centering
        \includegraphics[width=.95\linewidth]{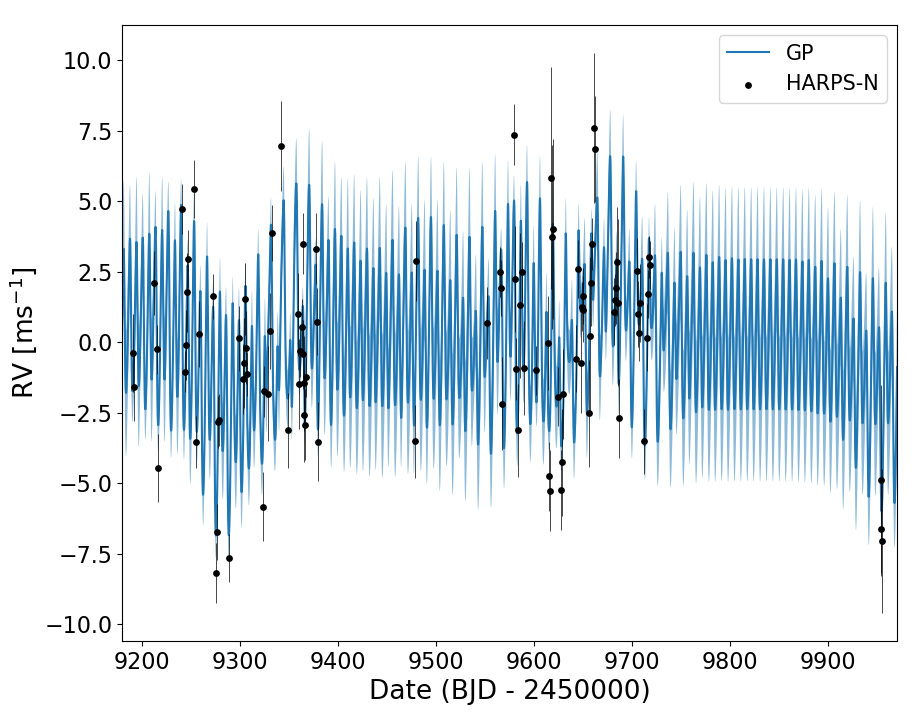}
    \end{minipage}
    \caption{Combined fit results obtained from \texttt{PyORBIT}. \textit{Left}: The RV fit to the HARPS-N s-BART reduced data. Black circular points show the phase folded HARPS-N RVs, grey points show these same values in subsequent phases, and the blue line shows the best fit Keplerian model. \textit{Right}: The RV and GP phase fit to the HARPS-N s-BART reduced data. Black round points show the HARPS-N RVs and the blue lines shows the corresponding GP fit to this data.}
    \label{fig:RV_fit}
\end{figure*}

Figure \ref{fig:transit_fit} shows the phase-folded light curves for both the \textit{TESS} and \textit{CHEOPS} data along with the best-fitting transit and GP models for the fit with the s-BART RVs. The left panel of Figure \ref{fig:RV_fit} shows the orbital solution and RV residuals phased on the period of HD 77946 b. The right panel of Figure \ref{fig:RV_fit} shows the s-BART HARPS-N (filled circles) RVs, together with the corresponding best-fit model of the planetary signal and the GP model of suspected stellar noise (blue line) as well as its associated uncertainty range (light blue shaded region).

We ensured that the nested sampling was converging to consistent results by running the same model multiple times, and across different machines. Given the aforementioned problems in constraining the stellar rotation period, we also ran models with different priors on $P_{\rm rot}$, despite this the planetary parameter results were still consistent to 1$\sigma$ on all runs. A corner plot with all the planetary parameters can be found in Appendix \ref{sec:appendix_fig}, see Figure \ref{fig:b_corners}. In addition to this the RV semi-amplitude (2.69 ± 0.57 ms$^{-1}$) obtained from the {\sc scalpels} analysis in Section \ref{sec:stellaractivity} was combined with the stellar mass and orbital inclination (1.17 M$_{\odot}$ and 87.10 degrees respectively) which lead to a mass determination of 8.83 ± 1.86 M$_{\oplus}$. This result is consistent with the results from the joint GP, photometric, and spectroscopic fit determined using \texttt{PyORBIT}. Fits were also performed with and without a GP, and comparing the Bayesian Information Criterion (BIC), the non-GP fit had a log-median BIC of -64134 whilst the fit with a GP had a log-median BIC of -56987. \citet{Trotta_2008} suggests under the assumption of a multivariate Gaussian posterior that a log-Bayes factor of $> 5$ should be considered "strong" evidence for a model. In this case the log-Bayes factor is 3573.5 ($\frac{1}{2}\Delta$BIC), further justifying the use of a GP.

In order to further confirm this, a phase folded plot of the HARPS-N RVs colour coded to the time period they were taken was produced (see Appendix \ref{sec:appendix_fig} Figure \ref{fig:rv_phase_colour}). From the grouping of RVs in this plot, we were able to conclude that the results were not being skewed by only sampling some parts of the orbital phase. An additional check was performed to ensure the validity of the eccentricity obtained, in which a 1-planet fit was attempted where the planets orbit was forced to be circular (with $e=0$). From this we calculate a slightly lower mass, $M_{\rm b}=7.68_{-1.25}^{+1.29}$\,M$_{\oplus}$, corresponding to a similar confidence detection of $\sim6\sigma$, with $R_{\rm b}=2.83_{-0.07}^{+0.07}$\,R$_{\oplus}$. Comparing the Bayesian Information Criterion (BIC), the forced circular orbit had a log-median BIC of -57009 whilst the eccentric orbit had a log-median BIC of -56987. Thus, whilst the confidence detections on $M_{\rm b}$ are both around $\sim6\sigma$, the median BIC of the eccentric orbit is better. In addition to this, the log-Bayes factor is 11 ($\frac{1}{2}\Delta$BIC), which supports choosing the more eccentric model. Orbits for single-planet systems are also not typically zero \citep{VanEylen2019} and there is no compelling argument to fix the eccentricity of this planet as circular, therefore we favour the model with an eccentric orbit.

The RV time series shown in Figure \ref{fig:RV_plot} also has a hint of a longer term trend.  However, as illustrated in Figure \ref{fig:RV_fit}, this is captured by the GP in our aforementioned fits as we did not fit for a linear trend. We did carry out some additional model runs allowing for a longer period Keplerian, in addition to the known Keplerian (HD 77946 b) and a GP. However, these failed to converge on a longer-period Keplerian signal. Whilst this does not mean that we can rule out the presence of a longer-period companion, the existing data does not provide support for such a signal. Further RV observations with HARPS-N are ongoing which will allow us to further investigate the possibility of a longer-period companion, as the existence of a gas giant in the system could determine the internal composition of the inner sub-Neptune by pebble filtering, modulating the water mass fraction of the planet \citep{Bitsch2021}.




\section{Discussion}
\label{sec:discussion}

\begin{figure*}
    \centering
    \includegraphics[width=0.8\linewidth]{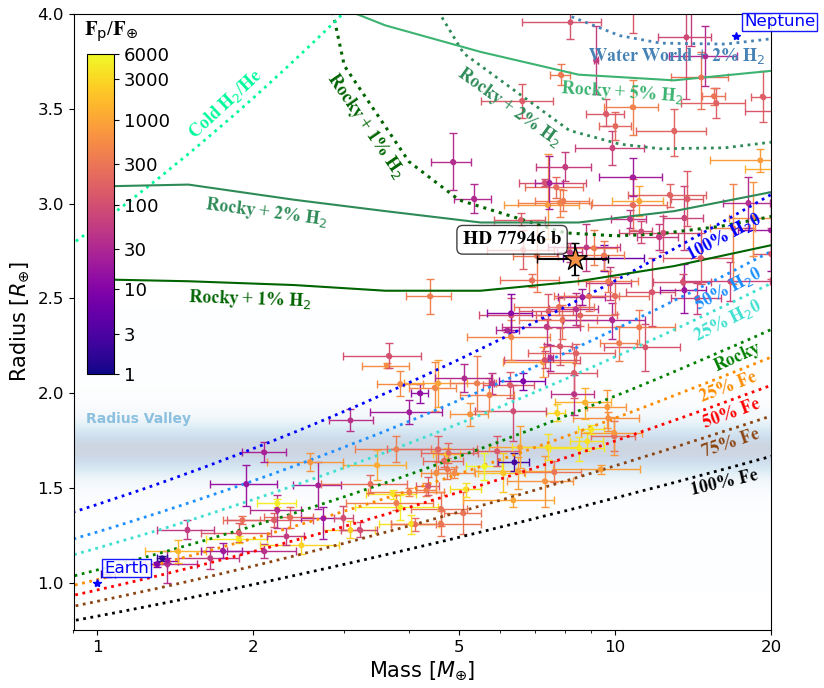}
    \caption{M-R diagram for HD 77946 b. HD 77946 b is indicated by the star and other confirmed planets with masses and radii measured to be better than 20\% and 10\% respectively, taken from the Extrasolar Planets Encyclopaedia, are also shown. The dotted lines show different planet compositions, taken from \citet{zeng_growth_2019} (1000K), and the solid lines show compositions taken from \citet{Lopez_2014} (1 Gyr, solar metallicity and flux of 1000F$_{\oplus}$). The data points are colour-coded according to their incident flux. Also shown are Earth and Neptune, for reference, and we indicate the approximate location of the radius valley \citep{Fulton_2017}.}
    \label{fig:radiusvalley_plots}
\end{figure*}

Our analysis allowed us to estimate the mass of HD 77946 b with a precision better than 20\%, and the radius with a precision better than 10\%. Combining the modelled mass and radius, we find the planet density to be $\rho_{\rm b} = 2.32_{-0.42}^{+0.44}$ g cm$^{-3}$. We place the mass and radius measurements of HD 77946 b into context and compare them to composition models in Figure \ref{fig:radiusvalley_plots}, which shows an M-R plot for small planets ($R < 4$ $R_{\oplus}$ and $M <$ 20 $M_{\oplus}$). Confirmed planets, with a mass and radius estimated to a precision better than 20\% and 10\% respectively, taken from the Extrasolar Planets Encyclopedia\footnote{\url{http://www.exoplanet.eu}} are also shown. Furthermore, the coloured lines in the M-R plot show different compositions, taken from
\citet{zeng_growth_2019}\footnote{Models are available online at \url{https://lweb.cfa.harvard.edu/~lzeng/planetmodels.html}} (dashed lines) and \citet{Lopez_2014} (solid lines). We include both models due to the recent conclusion of \citet{Rogers2023} in which it is found that the H/He mass relations in \citet{zeng_growth_2019} are not applicable to planets undergoing atmospheric evolution. There are a range of \citet{zeng_growth_2019} models available dependent on temperature, often incorrectly selected based on the equilibrium temperature of the planet. These models assume constant specific entropy, so should instead be selected based on the temperature of the corresponding specific entropy at 100-bar level in the gas envelope. We include the \citet{zeng_growth_2019} models at 1000K (based on HD 77946 b's equilibrium temperature calculated in Section \ref{sec:interior_comp}) in Figure \ref{fig:radiusvalley_plots} so that a direct comparison can be made to previous mass-radius studies in the literature, but note that the \citet{Lopez_2014} models are preferred as the assumption of a constant planet age is made instead of constant specific entropy, which is more suitable for planets undergoing atmospheric mass-loss. As can be seen in Figure \ref{fig:radiusvalley_plots}, HD 77946 b is consistent with having a sub-Neptune composition with a H/He atmosphere making up about $\sim$1\% of its mass or with having a water-world composition with a steam atmosphere.


\subsection{Interior Structure}
\label{sec:interior_comp}

To estimate the internal structure of the planet, we used a Bayesian analysis following the method described in \citet{Dorn2015,Dorn2017}, the same method employed to analyze the systems L98-59 \citep{Demangeon2021}, TOI-178 \citep{Leleu2021}, TOI-1064 \citep{Wilson2022}, and $\nu^2$ Lupi \citep{Delrez2021}. Our model assumes that the planet has four layers: an inner core made of iron and sulfur, a mantle of silicate (Si, Mg and Fe), a water layer and a gaseous layer made of pure H-He. The equations of state used are from \citet{Hakim2018} (iron core), \citet{Haldemann2020} (water layer) and \citet{Sotin2007} (silicate mantle). These three layers constitute the ‘solid’ part of the planet. The thickness of the pure H-He layer depends on stellar age and irradiation as well as the mass and radius of the solid part \citep{Lopez_2014}. 

Our Bayesian model fits the stellar and planetary observed properties to derive the posterior distributions of the internal structure parameters. The planetary parameters modeled are the mass fractions of each layer, the iron molar fraction in the core, the silicon and magnesium molar fraction in the mantle, the equilibrium temperature, and the age of the planet (equal to the age of the star). Uniform priors are used for these parameters, except for the mass of the gas layer which is assumed to follow a uniform-in-log prior, with the water mass fraction having an upper boundary of 0.5 \citep{Thiabaud2014,Marboeuf2014}. For more details related to the connection between observed data and derived parameters, we refer to \citet{Leleu2021}.  Finally, we assumed the Si/Mg/Fe molar ratios of each planet to be equal to the stellar atmospheric values.
We note that, as in many Bayesian analyses, the results (presented below) depend to some extent on the choice of the priors, which we selected in accordance with \citet{Dorn2017,Dorn2018} and \citet{Leleu2021}.

\begin{figure}
		\centering
		\includegraphics[width=\linewidth]{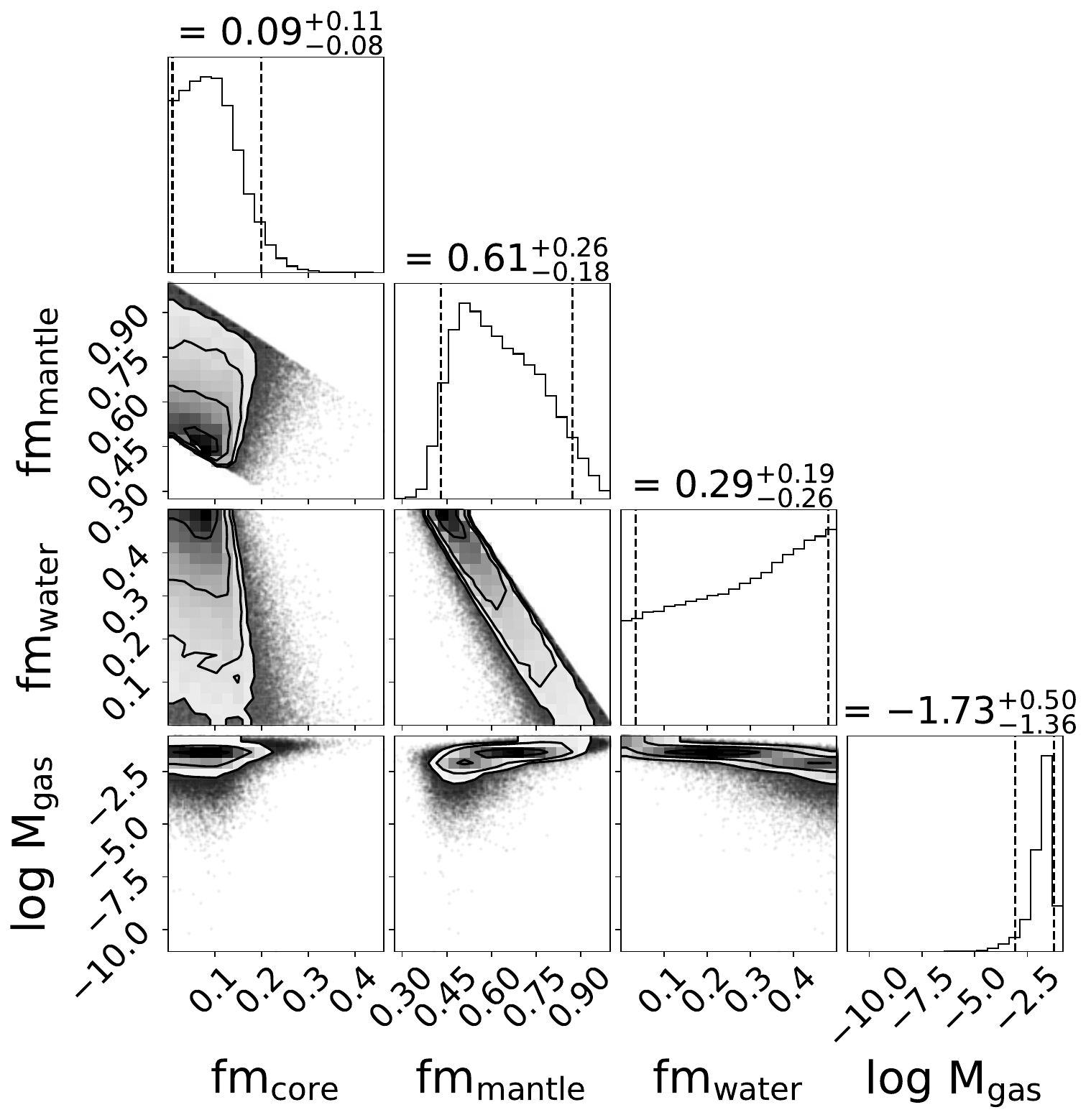}  
		\caption{Posterior distributions of the main parameters describing the internal structure of HD 77946 b. The corner plot shows the mass fraction of the inner core and of the water layer, the molar fractions of silicon and magnesium in the mantle, the iron molar fraction in the inner core, and the mass of gas in logarithmic scale. On top of each column are printed the mean and the 5 percent and 95 percent quantiles values.}
		\label{corner_plot}    
\end{figure}

The result of the internal structure modeling for HD 77946 b is shown in Figure \ref{corner_plot}. The planet shows a relatively small gaseous envelope for a Sub-Neptune of 0.02$^{+0.04}_{-0.02}$ M$_{\oplus}$ corresponding to less than 1\% of its total mass. However, due to its very high equilibrium temperature (1248$^{+40}_{-38}$ K), the radius of the gas layer is 0.69$^{+0.33}_{-0.33}$ R$_{\oplus}$, which corresponds to 1/4 of the HD 77946 b's total radius. Compared to the Earth, thanks to a possibly thick water layer ($0.29_{-0.26}^{+0.19}$), the iron core is proportionally smaller ($0.09_{0.08}^{+0.11}$ mass fraction in HD 77946 b compared to it being $\sim$1/3 of the Earth’s mass).

The amount of water on the planet is almost unconstrained ($0.29^{+0.19}_{-0.26} M_{\rm solid}$) due to the presence of an atmosphere. As the planet is close to its star we could assume that the planet formed inside the ice-line, that is to say it is a water-poor planet. We tried to model this eventuality by constraining the maximum fraction of water to $10^{-3}$. In this case, the envelope mass is bigger ($0.06^{+0.04}_{-0.03}$ M$_{\oplus}$) in order to retrieve a similar planet density, as well as its radius ($0.96^{+0.24}_{-0.24}$), with the same $M_{\rm core}/M_{\rm mantle}$. Both scenarios are plausible, however we favor the water-rich environment because the water-poor environment requires a strong assumption. Although giving slightly different results, they are both able to describe a planet of this mass and this radius. 

\subsection{Compositional confusion surrounding sub-Neptunes}
\label{sec: comp_confusion}

As seen in Figure \ref{fig:radiusvalley_plots}, HD 77946 b resides in an area on the M-R diagram where a degeneracy exists between the water-world and silicate/iron-hydrogen models. Whilst there have been strong arguments in favour of the existence of water-worlds \citep{zeng_growth_2019}, they cannot be identified based on the mass-radius relationship alone unless we can rule out a significant gas layer \citep{Adams2008}. Our interior structure modelling of HD 77946 b in Section \ref{sec:interior_comp} concludes that there is a H/He atmosphere that makes up a reasonable fraction of the radius, and the inclusion of both \citet{zeng_growth_2019} and \citet{Lopez_2014} models in Figure \ref{fig:radiusvalley_plots} reinforces this. However, the planet is very hot ($T_{\rm eq}\sim1250$~K) and as this is a highly irradiated planet water may be in supercritical state as shown in \citet{Mousis2020}. The interior structure modelling in Section \ref{sec:interior_comp} does not take into account different states of water, but even in the case of a solid layer of ice water our model still output a high water mass fraction of $30_{-26}^{+18}$\%. \citet{Aguichine2021} shows that planets with a water mass fraction between 20-50\% and irradiation temperatures of $\sim1000$K could also explain the location of HD 77946 b in the M-R diagram.

With an equilibrium temperature of $\sim$1250 K, it can be seen from Figure \ref{fig:radiusvalley_plots} that HD 77946 b is one of the hottest planets in this region of the M-R diagram. This would indicate that if HD 77946 b has a H/He atmosphere then it is escaping. Following the mass and radius predictions of \citet{Kubyshkina2022}, HD 77946 b resides roughly on the upper boundary of the 1350-1700K "migrated scenario", an escape-dominated region of M-R plots. This would indicate that for a H/He atmosphere HD 77946 b's evolution is mainly shaped by atmospheric escape and that atmospheric loss, due to the planets high thermal energy and low gravity throughout the first few tens of megayears of its evolution, is the dominant mechanism in driving the atmospheric evolution of the planet. Whilst the \citet{Lopez_2014} grids establish that a 1\% H/He atmosphere matches the structure at the planet's current-day size, mass, and irradiation, these models do not include photoevaporation. However, photoevaporation models do predict some sub-Neptunes above the valley at 6 days \citep{Owen2017}, and although these are on the hotter end of the sub-Neptunes, they are actually expected from population-level photoevaporation models. In addition to this, using the ATES grid from \citet{Caldiroli2022} the predicted lifetime for a 1\% H/He atmosphere is 1-3 Gyr. This makes a 1\% H/He atmospheric composition entirely reasonable given our estimated age of the system of 3.5 Gyr. In this case, the primordial parameters characterising the planet play a minor role in HD 77946 b's location in Figure \ref{fig:radiusvalley_plots}, as a wide range of possible initial parameters could have lead to the same outcome \citep{Kubyshkina2018, Kubyshkina2019, Kubyshkina2020}. Planets such as this, often called sub-Neptunes, sit above the radius valley, whilst stripped-core planets which are significantly smaller in size, often called super-Earths, sit below. Further studying larger sub-Neptunes, like HD 77946 b, will be key to constraining fundamental planetary formation and evolution processes such as atmospheric accretion and escape, migration, and core-composition. Furthermore, the precision of HD 77946 b's mass measurement means that this key parameter can be considered when endeavoring to understand these planetary pathways, as currently the hydrostatic upper atmospheres of small, low-irradiated exoplanets is mostly unexplored except for the Jeans escape model \citep{Konatham2020}.


\subsection{Prospects for atmospheric observations}
\label{sec:atmos_obs}

\begin{figure}
		\centering
		\includegraphics[width=\linewidth]{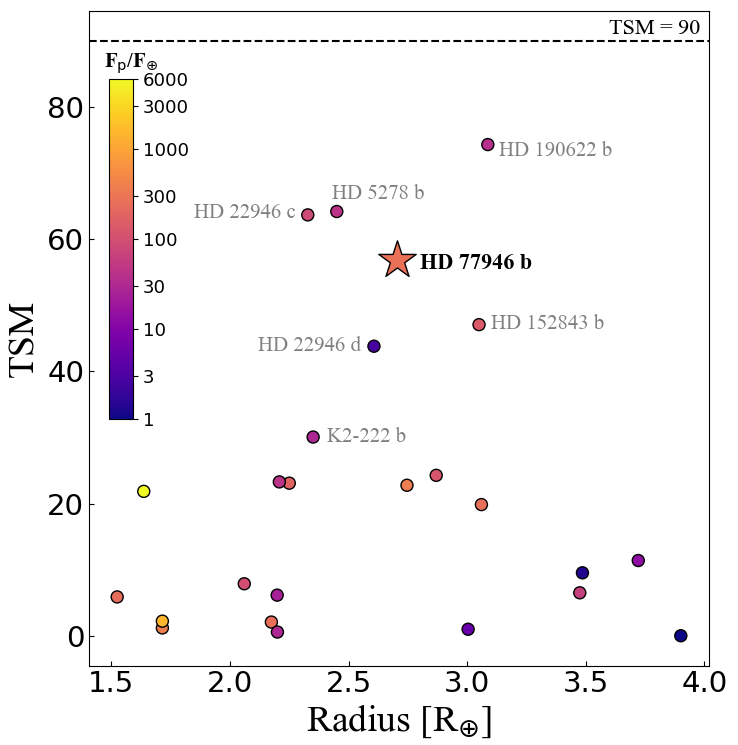}  
		\caption{Plot of planetary radius vs TSM for host stars with an effective temperature greater than 6000K. HD 77946b is indicated by the star and other confirmed planets, taken from the Extrasolar Planets Encyclopedia, are also shown colour-coded by their incident flux.} Additionally plotted is the recommended TSM$=90$ threshold (horizontal dashed-line). Whilst there are currently no sub-Neptunes orbiting hot stars that pass this recommended threshold, we note that HD 77946 b is one of the top 5 candidates in this area.
		\label{fig:TSM_plot}    
\end{figure}

We calculated the transmission spectroscopic metric (TSM) and the emission spectroscopic metric (ESM) outlined in \citet{Kempton_2018}, for follow-up atmospheric observations with \textit{JWST}, but find that the TSM and ESM (56.81 and 5.63 respectively) do not meet the threshold presented (suggested threshold of TSM > 90 for 1.5 < R$_{\rm b}$ < 10 R$_\oplus$, and ESM > 7.5). Even though HD 77946 b may not have a high enough TSM, or ESM, to be amongst the ‘best-in-class’ targets highlighted by \citet{Hord2023}, it still has one of the highest TSMs for a sub-Neptune orbiting a hot star ($T_{\rm eff} > 6000$K), as shown in Figure \ref{fig:TSM_plot}, so is still an interesting target for transmission spectroscopy. However, the brightness of the target, at K=7.6, would saturate many instrument modes on \textit{JWST}, but we emphasise that whilst this target may not be suitable for \textit{JWST} spectroscopy, a single transit should give a clear detection of the atmosphere if no clouds or hazes were present. We verified this with PandExo models with equilibrium chemistry and no clouds or hazes, and found that G140H (and potentially G235H) would be able to pick up water features in a H/He-dominated atmosphere without fully saturating the instrument, so studies of its atmosphere for other reasons than just spectrum investigations are also viable.

We also looked into the possibility of atmospheric He detection through transmission spectroscopy in the near-infrared. Scaling from \citet{Kirk2022}, who constrain the peak helium absorption for WASP-52b (J = 10.5, T$_{14} = 1.8\sim hr$) to 0.31\%, for HD 77946 (J = 7.9, T$_{14} = 3.3 \sim hr$) we could expect about 0.07\% precision with Keck/NIRSPEC on a similar signal for HD 77946. Thus, detection of a strong metastable He signal would be challenging but perhaps not impossible in multiple transits because of the bright star, given that \citet{Zhang2021} achieved better precision on 55 Cancri e. Still, some caution is warranted as our low log(R'$_{\rm HK}$) value indicates that HD 77946 does not exhibit strong chromospheric activity. Planets orbiting stars with log(R'$_{\rm HK}$) $\lesssim$ -5 are rarely detected to exhibit helium absorption \citep{Bennett2023}, and unfortunately, given the distance ($d=99$pc) and this inactivity, Lyman-$\alpha$ detection would be challenging due to interstellar absorption.

\subsection{Benefits of \textit{CHEOPS}}

The inclusion of photometric light curves from \textit{CHEOPS} improved the precision of our results significantly. In order to quantify this, the model was run a number of times with and without the various sets of data accrued. More specifically the model was run with only the HARPS-N s-BART RVs and sector 21 of the \textit{TESS} data, the HARPS-N s-BART RVs and both \textit{TESS} sectors, the HARPS-N s-BART RVs and sector 47 of the \textit{TESS} data alongside the \textit{CHEOPS} transit this sector overlaps with, the HARPS-N s-BART RVs with both \textit{TESS} sectors and the first \textit{CHEOPS} transit, and lastly the HARPS-N s-BART RVS with both \textit{TESS} sectors and both \textit{CHEOPS} transits. With the inclusion of the first \textit{CHEOPS} transit, the precision on the period and the transit centre time improved significantly (by 31\% and 43\% respectively), and with the inclusion of the second \textit{CHEOPS} transit the precision improved even further (by 37\% and 55\% respectively from the \textit{TESS} only values).
The precision on these parameters ensures future studies of this planet can be made without much time loss, and whilst this is not yet possible with the instrumentation on \textit{JWST}, in the future there is potential for this with more sensitive instrumentation. For instance, to plan observations of a transit of HD 77946 in a decade using the precision on the parameters from the model run with only \textit{TESS} photometric data, the uncertainty on the timing would be $\sim$28 mins. However, adding the \textit{CHEOPS} photometric data to the model means that this uncertainty reduces to only $\sim$17 mins. In addition to the aforementioned parameters, the precision on the planetary radius marginally improves (by 7\%) with the inclusion of the \textit{CHEOPS} data, which further highlights the benefits of the ultra-high precision photometry.





\section{Conclusions}
\label{sec:conclusion}

In this paper, we present the analysis of the \textit{TESS} target HD 77946, a bright (V = 9.00) F5 star (M$_\star$ = 1.17 M$_\odot$, R$_\star$ = 1.31 R$_\odot$). Using the results of a HARPS-N GTO campaign, as well as 2 additional \textit{CHEOPS} transits, alongside the 2 initial \textit{TESS} campaigns (each lasting around 27 days), we find HD 77946 to host one small (2.7 R$_\oplus$, 8.4 M$_\oplus$) transiting sub-Neptune, with an orbital period of $\sim$6.53 days. The accuracy of our mass measurement for this planet allows us to constrain likely composition scenarios. The planet is located above the radius valley, indicating that it is a sub-Neptune, and since the radius valley is theorised to separate super-Earths and sub-Neptunes the composition of the planet can be theorised in this context. We furthermore compare the properties of HD 77946's planet alongside those with similarly constrained radii and masses, and find that these suggest that HD 77946 b is a sub-Neptune with an iron-core fraction smaller than that of the Earth and an H/He atmosphere that is about $\sim$1\% of the total mass. 

This result contributes to the highlighted degeneracy between water-worlds and silicate/iron-hydrogen models. In the case of HD 77946 b the concept of a water-world remains ambiguous, whilst both \citet{zeng_growth_2019} and \citet{Lopez_2014} models suggest the existence of a $\sim$1\% H/He atmosphere, and our own internal composition models indicate that this H/He atmosphere makes up a significant proportion of the radius of the planet, the temperature of the planet means that that water could be in a supercritical state and so a steam atmosphere may be possible. This precise characterization of HD 77946 b makes it a valuable target for follow-up atmospheric observations to further explore this, as there are not many precisely characterized small planets amenable for atmospheric characterization orbiting stars like the Sun or bigger. Confirming or refuting the existence of water worlds for solar-type stars could have strong implications for our understanding of \textit{Kepler} demographics and our own solar system. This is also key to gain a better understanding of planetary compositions, which can in turn better inform the processes involved in planetary formation and evolution.

The combination of \textit{TESS}, \textit{CHEOPS}, and HARPS-N data in the characterisation of this planet, presents a promising approach in the use of multiple data sets from a range of telescopes in the improvement of precision of planetary properties. The addition of multiple transits by two space telescopes combined with 100+ RVs from a ground-based telescope, helped to significantly improve the radius and mass precision, placing it among the top ten mass and radius precision for confirmed F star planetary systems on the Extrasolar Planets Encyclopedia. This is promising for the future, as good sampling of the exoplanet population is needed to further inform planet formation models.

\section*{Acknowledgements}
The HARPS-N project was funded by the Prodex Program of the Swiss Space Office (SSO), the Harvard University Origin of Life Initiative (HUOLI), the Scottish Universities Physics Alliance (SUPA), the University of Geneva, the Smithsonian Astrophysical Observatory (SAO), the Italian National Astrophysical Institute (INAF), University of St. Andrews, Queen’s University Belfast, and University of Edinburgh.

We thank Suzanne Aigrain for her work in obtaining observations.

MPi acknowledges the financial support from the ASI-INAF Addendum n.2018-24-HH.1-2022 ``Partecipazione italiana al Gaia DPAC  - Operazioni e attivit\`a di analisi dati''. 

ACC and TW acknowledge support from STFC consolidated grant numbers ST/R000824/1 and ST/V000861/1, and UKSA grant number ST/R003203/1. KR acknowledges support from STFC Consolidated grant number ST/V000594/1. 

AAJ acknowledges support from a World-Leading St Andrews Doctoral Scholarship.

This work has been carried out within the framework of the NCCR PlanetS supported by the Swiss National Science Foundation under grants 51NF40\_182901 and 51NF40\_205606. 

R.D.H. is funded by the UK Science and Technology Facilities Council (STFC)'s Ernest Rutherford Fellowship (grant number ST/V004735/1).

FPE would like to acknowledge the Swiss National Science Foundation (SNSF) for supporting research with HARPS-N through the SNSF grants nr. 140649, 152721, 166227 and 184618. The HARPS-N Instrument Project was partially funded through the Swiss ESA-PRODEX Programme.

This work has been carried out within the framework of the NCCR PlanetS supported by the Swiss National Science Foundation under grants 51NF40\_182901 and 51NF40\_205606.
This project has received funding from the European Research Council (ERC) under the European Union’s Horizon 2020 research and innovation programme (grant agreement SCORE No 851555).

This research has made use of data obtained from or tools provided by the portal exoplanet.eu of The Extrasolar Planets Encyclopaedia.

\section*{Data Availability}

This paper includes raw data collected by the \textit{TESS} mission, which are publicly available from the Mikulski Archive for Space Telescopes (MAST, \url{https://archive.stsci.edu/tess}). Raw data collected by \textit{CHEOPS} can be found using the file keys in Table \ref{tab:cheops_obs} at \url{https://cheops-archive.astro.unige.ch/archive_browser/}. Observations made with HARPS-N on the Telescopio Nazionale Galileo 3.6m telescope are available in Appendix \ref{sec:appendix_tables}.




\bibliographystyle{mnras}
\bibliography{main, comp_bib} 



\appendix

\section{Additional Figures}
\label{sec:appendix_fig}

\begin{figure}
    \centering
    \includegraphics[width=.9\linewidth]{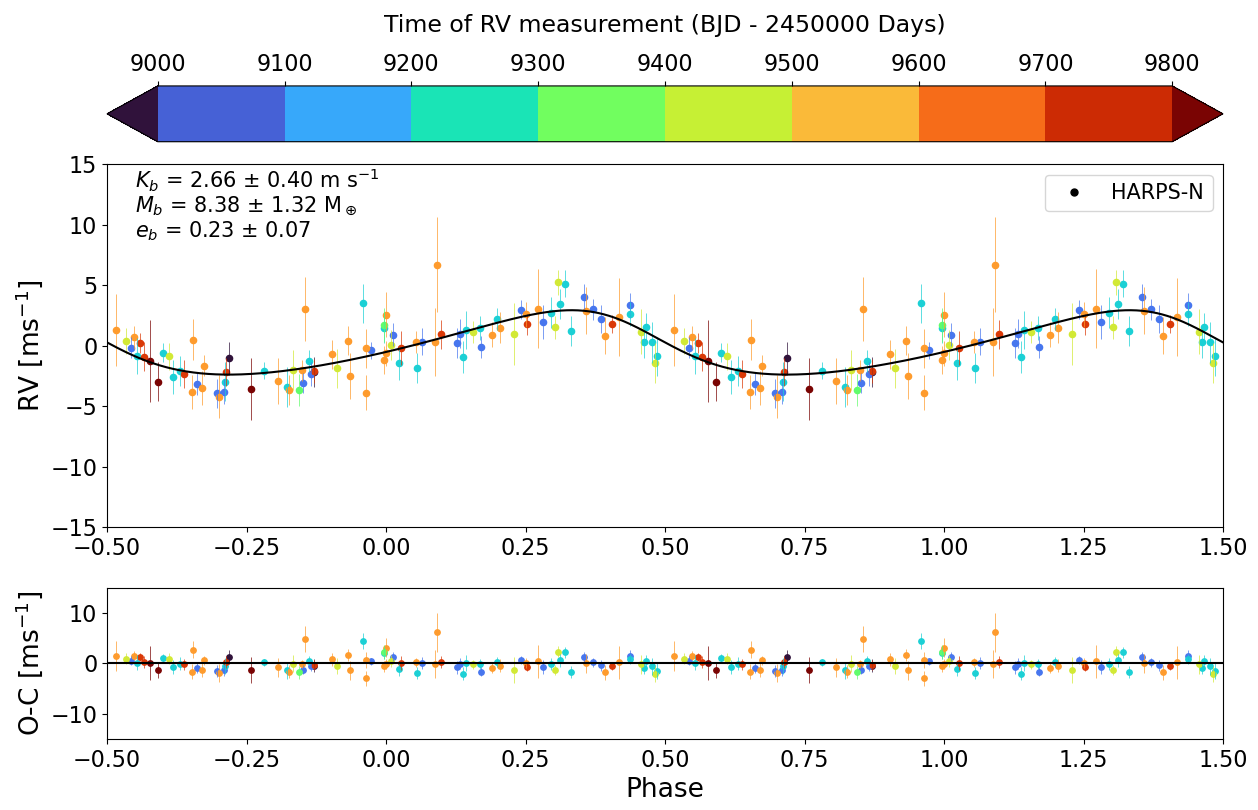}
    \caption{The RV fit to the s-BART reduced HARPS-N data colour coded by the time period the observation was taken. The black line shows the best fit Keplerian model.}
    \label{fig:rv_phase_colour}
\end{figure}

\begin{figure*}
    \centering
    \includegraphics[width=\textwidth]{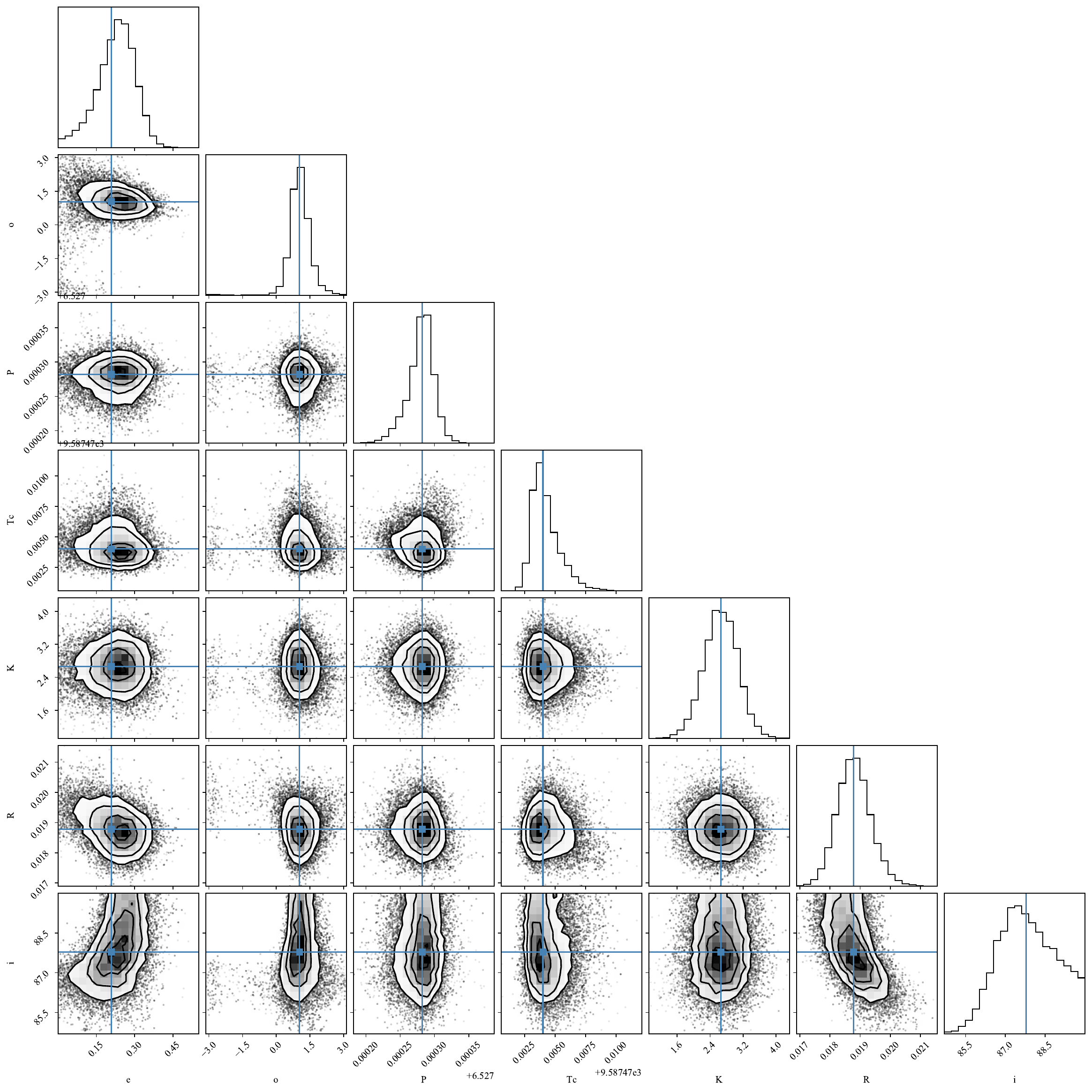}
    \caption{Corner plot of the best-fit planetary parameters obtained from the combined transit and s-BART RV fit.}
    \label{fig:b_corners}
\end{figure*}

\section{RV Data Tables}
\label{sec:appendix_tables}
\begin{table*}
    \centering
    \caption{HARPS-N radial velocity data and activity indicators (DRS 2.3.5).}
    \begin{tabular}{lllllllllll}
         \hline\hline 
         BJD\textsubscript{UTC} & RV & $\sigma$\textsubscript{RV} & BIS\textsubscript{span} & $\sigma$\textsubscript{BIS\textsubscript{span}} & FWHM & $\sigma$\textsubscript{FWHM} & H$\alpha$ & $\sigma$\textsubscript{H$\alpha$} & log$R'_{HK}$ & $\sigma$\textsubscript{log$R'_{HK}$}\\
         (d) & (m s\textsuperscript{-1}) & (m s\textsuperscript{-1}) & (m s\textsuperscript{-1}) & (m s\textsuperscript{-1}) & (km s\textsuperscript{-1}) & (km s\textsuperscript{-1}) &  & & &\\
         \hline
         \multicolumn{11}{l}{} \\
         2459190.738244 & 42545.63 & 1.43 & 22.64 & 2.88 & 7.89579 & 0.00288 & 0.190614 & 0.000198 & -5.00992 &	0.00243\\
         2459191.726875 & 42544.71 & 1.30 & 18.37 & 2.60 & 7.88572 & 0.00260 & 0.188756 & 0.000133 & -4.99872 & 0.00192\\
         2459212.579407 & 42547.40 & 1.16 & 21.49 & 2.32 & 7.88992 & 0.00232 & 0.187651 & 0.000108 & -5.00220 &	0.001670\\
         2459215.700791 & 42546.76 & 0.91 & 22.53 & 1.81 & 7.88838 & 0.00181 & 0.189548 & 0.000083 & -4.99492 &	0.001054\\
         2459216.701496 & 42542.55 & 1.27 & 16.67 & 2.54 & 7.88784 & 0.00254 & 0.192039 & 0.000152 & -5.01565 & 0.001934\\
         2459240.688690 & 42551.90 & 0.91 & 19.73 & 1.83 & 7.88491 & 0.00183 & 0.187665 & 0.000072 & -4.99512 & 0.001110\\
         2459244.615700 & 42545.98 & 0.74 & 18.82 & 1.49 & 7.88567 & 0.00149 & 0.191498 & 0.000082 & -5.00989 & 0.000746\\
         2459245.626303 & 42546.28 & 1.30 & 21.83 & 2.61 & 7.89188 & 0.00261 & 0.188959 & 0.000156 & -5.00007 & 0.001999\\
         2459246.628068 & 42548.00 & 1.47 & 20.14 & 2.94 & 7.88453 & 0.00294 & 0.188186 & 0.000176 & -5.02032 & 0.002631\\
         2459247.649564 & 42550.13 & 1.09 & 21.21 & 2.19 & 7.88688 & 0.00219 & 0.192860 & 0.000147 & -4.98634 & 0.001353\\
         2459253.634872 & 42551.17 & 1.06 & 22.19 & 2.12 & 7.88977 & 0.00212 & 0.185250 & 0.000089 & -5.00296 &	0.001462\\
         2459255.632035 & 42543.04 & 0.97 & 17.94 & 1.94 & 7.88397 & 0.00194 & 0.188543 & 0.000091 & -4.99380 & 0.001182\\
         2459258.707176 & 42548.39 & 1.19 & 14.96 & 2.38 & 7.88617 & 0.00238 & 0.188371 & 0.000131 & -5.02338 & 0.001902\\
         2459272.479370 & 42548.17 & 0.84 & 21.08 & 1.68 & 7.88774 & 0.00168 & 0.190507 & 0.000087 & -5.00828 & 0.000922\\
         2459275.535218 & 42539.73 & 1.10 & 22.52 & 2.20 & 7.89072 & 0.00220 & 0.189822 & 0.000128 & -5.01908 & 0.001554\\
         2459276.547869 & 42537.89 & 1.04 & 19.77 & 2.07 & 7.89502 & 0.00207 & 0.189724 & 0.000122 & -5.01309 & 0.001352\\
         2459277.511019 & 42542.46 & 0.98 & 23.18 & 1.97 & 7.88923 & 0.00197 & 0.189613 & 0.000115 & -4.99725 & 0.001206\\
         2459278.542373 & 42544.62 & 0.96 & 18.90 & 1.93 & 7.89235 & 0.00193 & 0.188118 & 0.000104 & -5.00032 & 0.001137\\
         2459289.510242 & 42538.21 & 0.88 & 23.75 & 1.75 & 7.88770 & 0.00175 & 0.187647 & 0.000083 & -4.99339 & 0.000967\\
         2459299.526078 & 42546.04 & 1.20 & 23.63 & 2.39 & 7.89351 & 0.00239 & 0.185450 & 0.000112 & -5.01840 & 0.001815\\
         2459303.509382 & 42543.54 & 1.31 & 23.79 & 2.61 & 7.88904 & 0.00261 & 0.187090 & 0.000118 & -5.00778 & 0.002149\\
         2459304.475734 & 42545.88 & 1.71 & 18.16 & 3.43 & 7.88886 & 0.00343 & 0.185810 & 0.000198 & -5.00684 & 0.003292\\
         2459305.468367 & 42546.94 & 1.34 & 21.28 & 2.67 & 7.89100 & 0.00267 & 0.186004 & 0.000128 & -4.99559 & 0.002097\\
         2459306.554217 & 42545.53 & 1.32 & 24.82 & 2.64 & 7.90420 & 0.00264 & 0.186187 & 0.000161 & -4.99972 & 0.002103\\
         2459307.460545 & 42545.04 & 0.82 & 21.70 & 1.65 & 7.89597 & 0.00165 & 0.186539 & 0.000062 & -4.98568 & 0.000862\\
         2459323.484332 & 42541.54 & 1.27 & 20.53 & 2.54 & 7.88360 & 0.00254 & 0.187299 & 0.000156 & -5.00107 & 0.001955\\
         2459324.415964 & 42544.93 & 0.95 & 18.37 & 1.90 & 7.88709 & 0.00190 & 0.189235 & 0.000104 & -4.99814 &	0.001124\\
         2459328.486578 & 42543.98 & 1.74 & 21.94 & 3.49 & 7.89026 & 0.00349 & 0.188016 & 0.000196 & -5.00685 & 0.00363\\
         2459330.548136 & 42546.86 & 1.35 & 18.70 & 2.70 & 7.89051 & 0.00270 & 0.182204 & 0.000132 & -4.99043 & 0.00235\\
         2459332.497322 & 42550.91 & 1.04 & 20.95 & 2.08 & 7.88630 & 0.00208 & 0.182855 & 0.000086 & -4.97477 & 0.00135\\
         2459342.427322 & 42551.28 & 1.72 & 26.78 & 3.43 & 7.88266 & 0.00343 & 0.182009 & 0.000137 & -5.02066 & 0.00396\\
         2459343.439367 & 42556.16 & 2.04 & 24.80 & 4.07 & 7.88516 & 0.00407 & 0.180045 & 0.000184 & -5.03492 & 0.00515\\
         2459344.463297 & 42557.58 & 1.35 & 21.65 & 2.70 & 7.89582 & 0.00270 & 0.178534 & 0.000104 & -5.00018 &	0.00248\\
         2459345.464621 & 42556.06 & 1.30 & 22.83 & 2.60 & 7.89100 & 0.00260 & 0.182092 & 0.000097 & -4.99065 &	0.00239\\
         2459349.385292 & 42542.09 & 1.37 & 17.52 & 2.75 & 7.89782 & 0.00275 & 0.184820 & 0.000144 & -4.97997 &	0.00220\\
         2459359.374662 & 42547.11 & 1.57 & 13.99 & 3.14 & 7.88737 & 0.00314 & 0.181396 & 0.000151 & -5.00339 &	0.00293\\
         2459360.401225 & 42544.94 & 1.66 & 19.12 & 3.31 & 7.88807 & 0.00331 & 0.186620 & 0.000156 & -5.04058 & 0.00362\\
         2459361.378805 & 42544.74 & 0.94 & 21.42 & 1.87 & 7.89241 & 0.00187 & 0.184807 & 0.000089 & -4.99773 &	0.00114\\
         2459363.379202 & 42549.07 & 1.01 & 18.51 & 2.03 & 7.88823 & 0.00203 & 0.189136 & 0.000115 & -4.98885 &	0.00129\\
         2459364.379400 & 42549.12 & 1.11 & 20.17 & 2.23 & 7.89459 & 0.00223 & 0.183737 & 0.000126 & -4.99346 &	0.00157\\
         2459364.445783 & 42545.19 & 1.26 & 19.49 & 2.52 & 7.88835 & 0.00252 & 0.182185 & 0.000138 & -5.00053 &	0.00213\\
         249365.3897683 & 42544.38 & 1.77 & 16.04 & 3.55 & 7.90140 & 0.00355 & 0.188554 & 0.000239 & -5.01815 &	0.00379\\
         2459365.452059 & 42543.30 & 1.69 & 20.25 & 3.38 & 7.89046 & 0.00338 & 0.187437 & 0.000201 & -5.02172 &	0.00386\\
         2459366.394436 & 42543.29 & 1.32 & 21.51 & 2.63 & 7.89258 & 0.00263 & 0.189100 & 0.000150 & -4.98859 & 0.00215\\
         2459367.380219 & 42545.28 & 0.74 & 22.26 & 1.48 & 7.89265 & 0.00148 & 0.190189 & 0.000078 & -4.99028 &	0.00072\\
         2459377.372865 & 42548.58 & 1.30 & 18.26 & 2.60 & 7.89079 & 0.00260 & 0.182648 & 0.000132 & -4.98786 &	0.00206\\
         2459378.374164 & 42546.63 & 1.21 & 19.16 & 2.42 & 7.89463 & 0.00242 & 0.181863 & 0.000124 & -4.98886 &	0.00183\\
         2459379.372684 & 42542.83 & 1.48 & 18.81 & 2.96 & 7.89083 & 0.00296 & 0.177918 & 0.000160 & -4.98670 &	0.00252\\
         2459478.754516 & 42542.05 & 1.35 & 21.39 & 2.70 & 7.89219 & 0.00270 & 0.191788 & 0.000141 & -4.99190 &	0.00226\\
         2459479.750471 & 42550.05 & 1.46 & 20.17 & 2.91 & 7.88647 & 0.00291 & 0.188824 & 0.000135 & -4.99578 &	0.00257\\
         2459551.637860 & 42540.76 & 1.34 & 20.88 & 2.68 & 7.87728 & 0.00268 & 0.195722 & 0.000200 & -4.96282 &	0.00220\\
         2459565.648685 & 42543.86 & 0.90 & 19.76 & 1.80 & 7.87775 & 0.00180 & 0.193419 & 0.000106 & -4.99928 &	0.00118\\
         2459566.603313 & 42543.09 & 1.02 & 17.98 & 2.05 & 7.87859 & 0.00205 & 0.195089 & 0.000125 & -4.98415 &	0.00147\\
         2459567.772083 & 42539.51 & 1.68 & 22.61 & 3.36 & 7.87449 & 0.00336 & 0.190469 & 0.000218 & -4.99283 &	0.00362\\
         2459579.695802 & 42549.13 & 1.09 & 19.57 & 2.19 & 7.88275 & 0.00219 & 0.193077 & 0.000128 & -4.97463 & 0.00162\\
         2459580.670788 & 42541.77 & 1.95 & 18.34 & 3.91 & 7.88148 & 0.00391 & 0.195502 & 0.000322 & -4.95991 &	0.00420\\
         2459581.666799 & 42538.54 & 1.10 & 22.15 & 2.20 & 7.88110 & 0.00220 & 0.195271 & 0.000133 & -4.96425 &	0.00162\\
         2459583.635824 & 42540.33 & 1.65 & 20.28 & 3.30 & 7.87606 & 0.00330 & 0.192377 & 0.000224 & -4.98113 &	0.00347\\
         2459585.703278 & 42542.86 & 2.68 & 23.96 & 5.36 & 8.87764 & 0.00536 & 0.194946 & 0.000395 & -5.06381 &	0.00925\\
         2459587.697808 & 42543.45 & 1.09 & 20.65 & 2.18 & 7.87787 & 0.00218 & 0.192123 & 0.000137 & -5.01491 &	0.00176\\
    \end{tabular}
    \label{tab:RV_HARPSN_1}
\end{table*}

\begin{table*}
    \centering
    \ContinuedFloat
    \caption{HARPS-N radial velocity data and activity indicators (DRS 2.3.5).}
    \begin{tabular}{lllllllllll}
         \hline\hline 
         BJD\textsubscript{UTC} & RV & $\sigma$\textsubscript{RV} & BIS\textsubscript{span} & $\sigma$\textsubscript{BIS\textsubscript{span}} & FWHM & $\sigma$\textsubscript{FWHM} & H$\alpha$ & $\sigma$\textsubscript{H$\alpha$} & log$R'_{HK}$ & $\sigma$\textsubscript{log$R'_{HK}$}\\
         (d) & (m s\textsuperscript{-1}) & (m s\textsuperscript{-1}) & (m s\textsuperscript{-1}) & (m s\textsuperscript{-1}) & (km s\textsuperscript{-1}) & (km s\textsuperscript{-1}) &  & & &\\
         \hline
         \multicolumn{11}{l}{} \\
         2459589.646447 & 42538.63 & 1.71 & 22.66 & 3.43 & 7.87687 & 0.00343 & 0.189918 & 0.000222 & -4.98076 & 0.00366\\
         2459601.661063 & 42539.88 & 0.88 & 20.03 & 1.75 & 7.87323 & 0.00175 & 0.191366 & 0.000082 & -4.97205 &	0.00107\\
         2459614.590547 & 42539.93 & 1.78 & 19.43 & 3.56 & 7.87874 & 0.00356 & 0.195972 & 0.000314 & -5.00648 & 0.00398\\
         2459615.707950 & 42536.50 & 1.26 & 24.59 & 2.52 & 7.87274 & 0.00252 & 0.194392 & 0.000186 & -4.98889 &	0.00220\\
         2459616.609675 & 42536.43 & 1.45 & 21.23 & 2.91 & 7.87942 & 0.00291 & 0.189195 & 0.000152 & -4.97938 &	0.00280\\
         2459617.446595 & 42544.01 & 4.30 & 32.76 & 8.41 & 7.88332 & 0.00841 & 0.184741 & 0.000621 & -5.32254 &	0.03471\\
         2459618.627553 & 42545.10 & 4.36 & 21.56 & 6.93 & 7.87238 & 0.00693 & 0.189146 & 0.000508 & -5.17112 & 0.01786\\
         2459619.568840 & 42541.27 & 3.42 & 8.780 & 6.84 & 7.86571 & 0.00684 & 0.187352 & 0.000499 & -5.10429 &	0.01492\\
         2459624.617176 & 42540.10 & 1.08 & 21.86 & 2.15 & 7.87520 & 0.00215 & 0.193279 & 0.000131 & -4.99474 &	0.00166\\
         2459627.630534 & 42536.75 & 1.43 & 20.21 & 2.86 & 7.88350 & 0.00286 & 0.191854 & 0.000191 & -4.99546 &	0.00276\\
         2459628.639713 & 42537.72 & 1.98 & 21.46 & 3.96 & 7.87955 & 0.00396 & 0.190605 & 0.000261 & -5.01533 &	0.00527\\
         2459629.667599 & 42541.17 & 1.59 & 28.67 & 3.19 & 7.88230 & 0.00319 & 0.188325 & 0.000178 & -5.00543 &	0.00362\\
         2459642.508152 & 42539.36 & 1.20 & 24.17 & 2.39 & 7.87693 & 0.00239 & 0.189119 & 0.000135 & -4.99967 &	0.00199\\
         2459644.596920 & 42544.26 & 1.05 & 20.74 & 2.09 & 7.87951 & 0.00209 & 0.190071 & 0.000104 & -4.98704 &	0.00155\\
         2459647.532098 & 42539.93 & 1.81 & 28.89 & 3.62 & 7.86876 & 0.00362 & 0.192875 & 0.000238 & -5.01207 & 0.00419\\
         2459648.495016 & 42542.60 & 0.88 & 19.67 & 1.76 & 7.88071 & 0.00176 & 0.189947 & 0.000082 & -4.95548 & 0.00105\\
         2459649.456712 & 42543.15 & 1.17 & 18.72 & 2.34 & 7.87702 & 0.00234 & 0.189671 & 0.000128 & -4.97015 & 0.00183\\
         2459649.485986 & 42542.42 & 1.00 & 19.22 & 2.00 & 7.88329 & 0.00200 & 0.193279 & 0.000115 & -4.96934 &	0.00137\\
         2459655.586609 & 42537.04 & 1.91 & 21.64 & 3.81 & 7.87221 & 0.00381 & 0.193593 & 0.000257 & -5.00888 &	0.00484\\
         2459656.579639 & 42537.05 & 2.90 & 21.05 & 5.80 & 7.87132 & 0.00580 & 0.191036 & 0.000430 & -5.10786 &	0.01208\\
         2459658.575254 & 42540.11 & 1.50 & 23.21 & 3.01 & 7.88011 & 0.00301 & 0.191272 & 0.000193 & -5.00224 &	0.00314\\
         2459659.593557 & 42544.49 & 0.92 & 18.34 & 1.84 & 7.87490 & 0.00184 & 0.188337 & 0.000115 & -4.98032 &	0.00123\\
         2459661.590946 & 42542.86 & 2.76 & 7.005 & 5.52 & 7.87246 & 0.00552 & 0.187074 & 0.000367 & -5.03862 &	0.00951\\
         2459662.543021 & 42544.09 & 1.92 & 17.34 & 3.84 & 7.87734 & 0.00384 & 0.187936 & 0.000236 & -5.00167 &	0.00465\\
         2459681.484604 & 42543.80 & 1.20 & 21.39 & 2.41 & 7.88442 & 0.00241 & 0.191549 & 0.000151 & -4.99980 &	0.00202\\
         2459682.470313 & 42544.09 & 0.94 & 18.47 & 1.88 & 7.87994 & 0.00188 & 0.186032 & 0.000083 & -4.96822 & 0.00122\\
         2459683.458239 & 42544.47 & 1.30 & 19.13 & 2.61 & 7.87934 & 0.00261 & 0.190440 & 0.000135 & -4.99794 &	0.00238\\
         2459684.457695 & 42543.91 & 1.99 & 23.69 & 3.97 & 7.88790 & 0.00397 & 0.186930 & 0.000261 & -5.03925 &	0.00544\\
         2459685.487443 & 42547.75 & 3.13 & 30.14 & 6.26 & 7.88418 & 0.00626 & 0.186686 & 0.000466 & -5.07349 &	0.01237\\
         2459686.495470 & 42540.80 & 1.43 & 24.38 & 2.86 & 7.87533 & 0.00286 & 0.189690 & 0.000173 & -4.99259 &	0.00277\\
         2459705.397453 & 42546.40 & 1.21 & 21.76 & 2.42 & 7.87375 & 0.00242 & 0.192274 & 0.000151 & -4.97071 &	0.00183\\
         2459706.358644 & 42543.39 & 1.19 & 18.78 & 2.38 & 7.88012 & 0.00238 & 0.191660 & 0.000151 & -4.97177 &	0.00187\\
         2459707.386372 & 42543.58 & 1.02 & 21.00 & 2.04 & 7.88241 & 0.00204 & 0.190483 & 0.000116 & -4.97150 &	0.00141\\
         2459708.408500 & 42543.11 & 1.39 & 19.23 & 2.78 & 7.87579 & 0.00278 & 0.189700 & 0.000158 & -4.99957 & 0.00272\\
         2459712.392219 & 42540.32 & 1.18 & 19.75 & 2.36 & 7.87731 & 0.00236 & 0.190551 & 0.000140 & -4.98207 &	0.00190\\
         2459715.401979 & 42541.48 & 1.23 & 20.88 & 2.47 & 7.88239 & 0.00247 & 0.195684 & 0.000197 & -5.04080 &	0.00234\\
         2459716.404532 & 42544.99 & 0.87 & 21.27 & 1.74 & 7.87171 & 0.00174 & 0.190347 & 0.000094 & -4.98060 &	0.00107\\
         2459717.404407 & 42545.85 & 0.75 & 18.86 & 1.51 & 7.87379 & 0.00151 & 0.188707 & 0.000067 & -4.98103 &	0.00083\\
         2459718.410118 & 42545.62 & 0.89 & 14.11 & 1.78 & 7.87559 & 0.00178 & 0.187069 & 0.000089 & -4.96388 &	0.00108\\
         2459953.507204 & 42535.22 & 3.63 & 26.95 & 7.26 & 7.87150 & 0.00726 & 0.192918 & 0.000584 & -5.16967 &	0.01880\\
         2459953.596313 & 42532.62 & 1.65 & 25.41 & 3.30 & 7.87617 & 0.00396 & 0.194549 & 0.000254 & -4.99257 &	0.00334\\
         2459954.680378 & 42533.91 & 2.66 & 22.81 &	5.32 & 7.86896 & 0.00532 & 0.190955 & 0.000396 & -4.98334 &	0.00724\\
         \hline
    \end{tabular}
    \label{tab:RV_HARPSN_2}
\end{table*}

\begin{table*}
    \centering
    \caption{HARPS-N radial velocity data (s-BART).}
    \begin{tabular}{lll}
        \hline\hline 
        BJD\textsubscript{UTC} & RV & $\sigma$\textsubscript{RV}\\
        (d) & (m s\textsuperscript{-1}) & (m s\textsuperscript{-1})\\
        \hline
        \multicolumn{3}{l}{} \\
        2459190.738244 & 42450.83 & 1.38\\
        2459191.726875 & 42449.65 & 1.23\\
        2459212.579407 & 42453.33 & 1.13\\
        2459215.700791 & 42450.97 & 0.87\\
        2459216.701496 & 42446.76 & 1.20\\
        2459240.688690 & 42455.95 & 0.87\\
        2459244.615700 & 42450.18 & 0.71\\
        2459245.626303 & 42451.12 & 1.25\\
        2459246.628068 & 42452.99 & 1.39\\
        2459247.649564 & 42454.18 & 1.03\\
        2459253.634872 & 42456.65 & 1.02\\
        2459255.632035 & 42447.69 & 0.94\\
        2459258.707176 & 42451.51 & 1.16\\
        2459272.479370 & 42452.85 & 0.81\\
        2459275.535218 & 42443.03 & 1.06\\
        2459276.547861 & 42444.50 & 0.99\\
        2459277.511019 & 42448.39 & 0.93\\
        2459278.542373 & 42448.47 & 0.92\\
        2459289.510242 & 42443.57 & 0.85\\
        2459299.526078 & 42451.37 & 1.16\\
        2459303.509382 & 42449.92 & 1.25\\
        2459304.475734 & 42450.49 & 1.61\\
        2459305.468367 & 42452.76 & 1.27\\
        2459306.554217 & 42451.03 & 1.30\\
        2459307.460545 & 42450.09 & 0.78\\
        2459323.484332 & 42445.39 & 1.22\\
        2459324.415963 & 42449.48 & 0.92\\
        2459328.486578 & 42449.40 & 1.68\\
        2459330.548136 & 42451.62 & 1.34\\
        2459332.497322 & 42455.09 & 1.01\\
        2459342.427323 & 42458.18 & 1.61\\
        2459343.439367 & 42458.79 & 1.90\\
        2459344.463297 & 42461.02 & 1.28\\
        2459345.464621 & 42461.64 & 1.23\\
        2459349.385292 & 42448.10 & 1.34\\
        2459359.374662 & 42452.21 & 1.48\\
        2459360.401225 & 42449.75 & 1.60\\
        2459361.378805 & 42450.90 & 0.91\\
        2459363.379202 & 42451.75 & 0.99\\
        2459364.379401 & 42454.72 & 1.09\\
        2459364.445783 & 42450.80 & 1.23\\
        2459365.389768 & 42449.77 & 1.68\\
        2459365.452059 & 42448.63 & 1.64\\
        2459366.394436 & 42448.30 & 1.26\\
        2459367.380219 & 42449.98 & 0.71\\
        2459377.372865 & 42454.54 & 1.26\\
        2459378.374164 & 42451.95 & 1.19\\
        2459379.372684 & 42447.69 & 1.40\\
        2459478.754516 & 42447.71 & 1.31\\
        2459479.750471 & 42454.10 & 1.42\\
        2459551.637860 & 42446.28 & 1.26\\
        2459565.648685 & 42448.10 & 0.89\\
        2459566.603313 & 42447.51 & 0.99\\
        2459567.772083 & 42443.43 & 1.66\\
        2459579.695802 & 42452.97 & 1.07\\
        2459580.670788 & 42447.86 & 1.88\\
        2459581.666799 & 42444.65 & 1.09\\
        2459583.635824 & 42442.48 & 1.64\\
        2459585.703278 & 42446.91 & 2.57\\
        2459587.697808 & 42448.09 & 1.08\\
    \end{tabular}
    \label{tab:RV_HARPSN_3}
\end{table*}

\begin{table*}
    \centering
    \ContinuedFloat
    \caption{HARPS-N radial velocity data (s-BART).}
    \begin{tabular}{lll}
        \hline\hline 
        BJD\textsubscript{UTC} & RV & $\sigma$\textsubscript{RV}\\
        (d) & (m s\textsuperscript{-1}) & (m s\textsuperscript{-1})\\
        \hline
        \multicolumn{3}{l}{}\\        
        2459589.646447 & 42444.70 & 1.67\\
        2459601.661063 & 42444.61 & 0.87\\
        2459614.590547 & 42445.58 & 1.68\\
        2459615.707950 & 42440.85 & 1.22\\
        2459616.609675 & 42440.34 & 1.43\\
        2459617.446595 & 42451.43 & 3.94\\
        2459618.627553 & 42449.33 & 3.27\\
        2459619.568840 & 42449.62 & 3.21\\
        2459624.617176 & 42443.68 & 1.06\\
        2459627.630534 & 42440.36 & 1.43\\
        2459628.639713 & 42441.35 & 1.90\\
        2459629.667599 & 42443.76 & 1.60\\
        2459642.508152 & 42445.03 & 1.18\\
        2459644.596920 & 42448.21 & 1.03\\
        2459647.532098 & 42444.88 & 1.77\\
        2459648.495016 & 42446.86 & 0.87\\
        2459649.456712 & 42446.76 & 1.16\\
        2459649.485986 & 42447.25 & 1.00\\
        2459655.586609 & 42443.09 & 1.89\\
        2459656.579639 & 42445.82 & 2.80\\
        2459658.575254 & 42447.69 & 1.51\\
        2459659.593557 & 42449.08 & 0.93\\
        2459661.590946 & 42453.21 & 2.66\\
        2459662.543021 & 42452.45 & 1.88\\
        2459681.484604 & 42446.69 & 1.20\\
        2459682.470313 & 42447.09 & 0.93\\
        2459683.458239 & 42447.54 & 1.28\\
        2459684.457695 & 42448.46 & 1.93\\
        2459685.487443 & 42447.01 & 2.96\\
        2459686.495470 & 42442.90 & 1.40\\
        2459705.397452 & 42448.11 & 1.20\\
        2459706.358644 & 42446.60 & 1.16\\
        2459707.386372 & 42445.93 & 1.00\\
        2459708.408500 & 42446.99 & 1.39\\
        2459712.392220 & 42442.09 & 1.16\\
        2459715.401979 & 42445.76 & 1.19\\
        2459716.404532 & 42447.31 & 0.86\\
        2459717.404407 & 42448.64 & 0.74\\
        2459718.410118 & 42448.34 & 0.86\\
        2459953.507204 & 42440.70 & 3.40\\
        2459953.596313 & 42438.99 & 1.58\\
        2459954.680378 & 42438.55 & 2.55\\
        \hline
    \end{tabular}
    \label{tab:RV_HARPSN_4}
\end{table*}


\bsp	
\label{lastpage}
\end{document}